\documentclass[aps,prd,amsmath,twocolumn,superscriptaddress]{revtex4}

\usepackage{hyperref}
\usepackage{ifthen}
\usepackage{amsmath}
\usepackage{amssymb}
\usepackage{color}
\usepackage{graphicx}
\usepackage{xfrac}

\newcommand*{\VEC}[1]  {\textbf{#1}}

\newcommand*{\df}  {\delta}

\newcommand*{\vp}  {\textbf{p}}
\newcommand*{\vk}  {\textbf{k}}
\newcommand*{\vq}  {\textbf{q}}

\newcommand*{\Pp} {\mathbb{P}}

\begin{document}

\title{Lagrangian perturbation theory at one loop order: successes, failures, and improvements}
\author{Zvonimir Vlah}
\email{zvlah@physik.uzh.ch}
\affiliation{Physik-Institut, University of Z\"{u}rich, Winterthurerstrasse 190, CH-8057, Z\"{u}rich, Switzerland}
\affiliation{Institute for Computational Science, University of Z\"{u}rich, Winterthurerstrasse 190, CH-8057, Z\"{u}rich, Switzerland}

\author{Uro$\check{\rm s}$ Seljak} 
\affiliation{Physics, Astronomy Department, University of California, Berkeley, California, USA.} 
\affiliation{Lawrence Berkeley National Laboratory, Berkeley, CA, USA}

\author{Tobias Baldauf}
\affiliation{Institute for Advanced Study, Princeton, NJ 08544, USA}

\begin{abstract}
We apply the convolved Lagrangian perturbation theory (CLPT) formalism, in which one can express the matter 
density power spectrum in terms of integrals over a function of cumulants of the displacement 
field, allowing for a resummation of the terms, to evaluate the full one loop power spectrum. 
We keep the cumulants up to third order, extending the Zel'dovich approximation and providing 
the power spectrum analogous to the calculations recently performed for the correlation function. We 
compare the results to the N-body simulations and to the Lagrangian perturbation simulations up to the second order. 
We find that the analytic calculations are in a good agreement with the LPT simulations, but
when compared to full N-body simulations, we find that while one loop calculations improve upon the Zel'dovich 
approximation in the power spectrum, they still significantly lack power. 
As found previously in the correlation function one loop CLPT improves slightly against Zel'dovich above 30Mpc/h, 
but is actually worse than Zel'dovich below that. 
We investigate the deficiencies of the CLPT approach and 
argue that main problem of CLPT is its inability to trap particles inside dark matter halos, which leads 
to an overestimate of the small scale power of the displacement field and
to an underestimate of the small scale power from one halo term effects. 
We model this using the displacement field damped at a nonlinear scale (CLPTs). 
To explore this in more detail we decompose 
the power spectrum and correlation function into three 
additive components: Zel'dovich, residual BAO wiggle, and residual broad band. 
One loop CLPT predicts small modifications to BAO wiggles that are  
enhanced in CLPTs, 
with up to 5\% corrections to correlation function around BAO scale. 
For the residual broad band contribution CLPTs 
improves the broad band power in power spectrum, but is still insufficient compared to simulations, 
and makes the correlation function agreement worse than CLPT.
\end{abstract}

\maketitle

\section{Introduction}
\label{sec:intro}

Clustering of dark matter particles under gravity represents one of the building blocs in the study 
of large scale structure (LSS). Understanding the non-linear effects of dark matter clustering is crucial 
for improving the theoretical modelling for many current cosmological probes like galaxy surveys, weak 
lensing etc. The current paradigm is that large-scale structure grows through a process of gravitational instability, 
starting from a nearly scale-invariant spectrum of Gaussian fluctuations at early times. Since dark matter 
particles are assumed to be non-relativistic, at scales smaller than the Hubble scale general relativistic 
description of gravity can be reduced to Newtonian description. On large scales (but inside the Hubble 
horizon) the matter distribution is well modelled by linear perturbation theory. Conversely, on small scales, 
or Fourier modes with $k>0.1$ Mpc$/h$, the dynamics starts to be non-linear. One way to address this 
are the numerical simulations of the N-body type which offer a reliable way to understand the nonlinear 
clustering of matter.

An alternative approach to the non-linear scales (at least in quasi linear regime) is to extend the perturbation 
theory beyond the linear order. Main advantages of this approach are twofold. From a practical side perturbation 
theory offers a faster way of evaluating the observables for a given set of cosmological parameters. These 
observables are then used for comparison with the measurements in order to put the constraints on cosmological 
parameters. From a theoretical perspective perturbation theory offers an additional and/or complementary 
physical insight into the effects of non-linear clustering. A better physical understanding would also be useful 
to model higher order correlations, such as modelling the covariance matrix of dark matter two-point correlations 
etc. 

Consequently, numerous approaches have been introduced for computing statistical properties of the matter 
distribution. The standard perturbation theory (SPT) in Eulerian framework has been extensively studied and 
has achieved some success (see for example \cite{Bouchet:1995ez, Bernardeau:2001qr, Carlson:2009it, 
Bernardeau:2013oda}). Various resummation schemes have been proposed \cite{Crocce:2005xy, Crocce:2005xz, 
Taruya:2007xy, Matarrese:2007wc, Bernardeau:2008fa, Bernardeau:2011dp, Bernardeau:2012ux, Anselmi:2012cn, 
Sugiyama:2013pwa} in order to extend the validity of the perturbative expansion. Numerical implementations 
of some of these methods have become available \cite{Crocce:2012fa, Taruya:2012ut}. Also, a number of 
alternative methods have been suggested (e.g. \cite{McDonald:2006hf, Pietroni:2008jx, Valageas:2006bi, 
Sugiyama:2012pc, Blas:2013aba}) that use different levels of approximation. Alternatively, one can also consider 
Lagrangian picture as starting point of Lagrangian perturbation theory (LPT), e.g. \cite{Matsubara:2007wj, Matsubara:2008wx, 
Okamura:2011nu, Carlson:2012bu, Valageas:2013gba, Wang:2013hwa, Valageas:2013hxa, Sugiyama:2013mpa, White:2014gfa}, 
where the focus is on perturbing the displacement field rather than overdensity and velocity fields itself. Recent 
work has emphasized the fundamental failure of ab initio perturbation theory on small scales, where effects are 
non-perturbative (e.g. \cite{Carrasco:2012cv, Carrasco:2013sva, Carrasco:2013mua, Pajer:2013jj, Mercolli:2013bsa, 
Carroll:2013oxa, Porto:2013qua, Senatore:2014via}). In this approach, called effective field theory of large scale structure (EFTofLSS),
small scales contributions are integrated out, and one is left with the effective theory formulation with free coefficients which 
are incorporating small scale contribution.

In this paper we first follow the recent work done in studying the Lagrangian picture in the context of LPT where 
the cumulants are kept in the exponent, CLPT 
\citep{Carlson:2012bu, Sugiyama:2013mpa}, extending the analytic calculation methods, and exploring the accuracy 
and performance of this approach for the matter power spectrum and correlation function. We test the performance 
of one loop analytical calculations against the N-body simulations in both  Fourier and configuration space. 
We then connect these calculations to the standard perturbation theory and show the connection between the two. 
We identify the main shortcomings of the approach and propose the decomposition of the power spectrum in 
tree additive parts; Zel'dovich part, residual contribution to the baryon acoustic oscillation (BAO) wiggles and 
residual contribution to the broad band power. We show that CLPT based approach is well suited for analysing the 
residual wiggle contribution. We show how the corrections of the displacement field of two point function
(see e.g. \cite{Chan:2013vao}), which we call CLPTs, affect the residual contributions to the BAO wiggles. For the residual broad band 
part we follow a similar approach investigating the effects of CLPTs on the power spectrum and correlation function.  
Finally, we show the relative effects of these contributions and comparison to the N-body simulations in both Fourier 
and configuration space.

This paper is organized as follows: in section \ref{sec:clustering} we present the framework for the dark matter power 
spectrum and review the Lagrangian perturbation theory for the displacement field.  We present the methods to compute 
the one loop power spectrum and show the corresponding low k limit result. In subsection \ref{sec:2LPTcross} we look at 
various cross-power spectra at 2LPT level and compare it to the grid 2LPT numerical results. In subsection \ref{sec:corrfnc} 
the correlation function results are presented and compared to the N-body measurements. In section \ref{sec:decomp} we 
study the improvement of the CLPT results by decomposing it into three additive parts and we show the extent of agreement 
of these results with N-body simulations on the power spectrum and the correlation function. Finally, we conclude our 
findings in section \ref{sec:conc}. In Appendices \ref{sec:XY}, \ref{sec:VT}, \ref{sec:int1}, \ref{sec:int2} we show some 
details of the calculations and write explicit forms of the terms contributing to the power spectra. 

For this work, flat $\Lambda$CDM model is assumed 
$\Omega_{\rm m}=0.272$, 
$\Omega_{\Lambda}=0.728$, $\Omega_{\rm b}/\Omega_{\rm m}=0.167$, $h=0.704$,
$n_s=0.967$, $\sigma_8=0.81$. The primordial density field is generated using the matter transfer
function by CAMB. The positions and velocities of all the dark matter
particles are given at the redshifts $z=0.0$, $0.5$, $1.0$, and $2.0$.

\section{Clustering in Lagrangian picture }
\label{sec:clustering}
\subsection{Overdensity field evolution and power spectrum}
\label{subsec:density}
A central quantity in Lagrangian picture is the displacement field $\Psi (\VEC{q},\tau)$. 
It represents the mapping of a particle from its initial position $\VEC{q}$, to the 
Eulerian-space coordinate at a given moment in time $\VEC{r}$
\begin{align}
 \VEC{r}(\VEC{q},\tau)=\VEC{q}+\Psi (\VEC{q},\tau).
 \label{eq:psidef}
\end{align}
From this we see that $\Psi (\VEC{q},\tau)$ can also be understood as the velocity 
field integral along the world-line of the particle, starting from the origin
\begin{align}
\Psi (\VEC{q},\tau)=\int^\tau d\tau'\VEC{v}\left(\VEC{r}(\VEC{q},\tau'),\tau'\right).
\end{align}
We are interested in the density field of dark matter particles and how it evolves with time. 
Continuity equation and the assumption that we have uniform initial density field give the relation of 
overdensity field in the volume element $d^3r$ at the position $\VEC{r}$ with initial conditions
\begin{align*}
&\left(1+\df(\VEC{r})\right)d^3r=d^3q\nonumber\\
&\qquad \to ~~1+\df(\VEC{r})=\int d^3q~\df^D\left(\VEC{r}-\VEC{q}-\Psi(\VEC{q})\right).
\end{align*}
In Fourier space this relation gives
\begin{align}
(2\pi)^3\df^D(\VEC{k})+\df(\VEC{k})=\int d^3 q ~ e^{i\VEC{k}\cdot\VEC{q}}~\exp{(i\VEC{k}\cdot\Psi)},
\label{eq:dfkspace}
\end{align}
where we are following the Fourier conventions:
\begin{align*}
    &\tilde{f}(\VEC{k})=\mathcal{F}\left[f(\VEC{x})\right](\VEC{k})=
    \int{ d^3x ~\text{exp}(i\VEC{k}\cdot\VEC{x})f(\VEC{x})},\nonumber\\
    &f(\VEC{x})=\mathcal{F}^{-1}\left[\tilde{f}(\VEC{k})\right](\VEC{x})=\int{\frac{d^3k}{(2\pi)^3}
            ~\text{exp}(-i\VEC{k}\cdot\VEC{x})\tilde{f}(\VEC{k})}.
\end{align*}
The simplest and thus the most interesting statistical quantity that can be constructed from
this field is a two point correlation function, or its Fourier space analog, the power spectrum. 
Since we assume a homogeneity and isotropy of the dark matter distribution 
we can define the power spectrum
\begin{align}
 (2\pi)^3P(k)\df^D(\VEC{k}+\VEC{k}')=\left\langle\df(\VEC{k})\df({\VEC{k}'})\right\rangle.
 \label{eq:PSdef}
\end{align}
Using equation \eqref{eq:dfkspace} it follows that the power spectrum in terms of displacement 
field is given by
\begin{align}
 (2\pi)^3\df^D(k)+P(k)=\int d^3 q ~e^{-i\VEC{q}\cdot\VEC{k}}\left\langle
 \exp(-i\VEC{k}\cdot\Delta)\right\rangle,
\label{eq:PSdis}
\end{align}
where we have introduced the differential displacement vector field
\begin{align}
 \Delta=\Psi(\VEC{q}_2)-\Psi(\VEC{q}_1)
\end{align}
and define the separation vector $\VEC{q}=\VEC{q}_2-\VEC{q}_1$. 
Following the notation from \cite{Carlson:2012bu} we can introduce the generating function 
of the differential displacement vector field
\begin{equation}
K(\VEC{q})=\left\langle\exp(-i\VEC{k}\cdot\Delta)\right\rangle,
\label{eq:defK}
\end{equation}  
As a consequence of spatial homogeneity and isotropy, generating function $K$ is a function 
of separation vector $\VEC{q}$ only, rather than $\VEC{q}_2$ and $\VEC{q}_1$. In this way the 
translational invariance remains manifestly imposed at every step of this approach.

\subsection{Cumulant expansion and the hierarchy}
\label{subsec:PT}

\begin{figure*}[tb]
   \begin{center}
   \hspace*{-0.5cm}
   \includegraphics[scale=0.49]{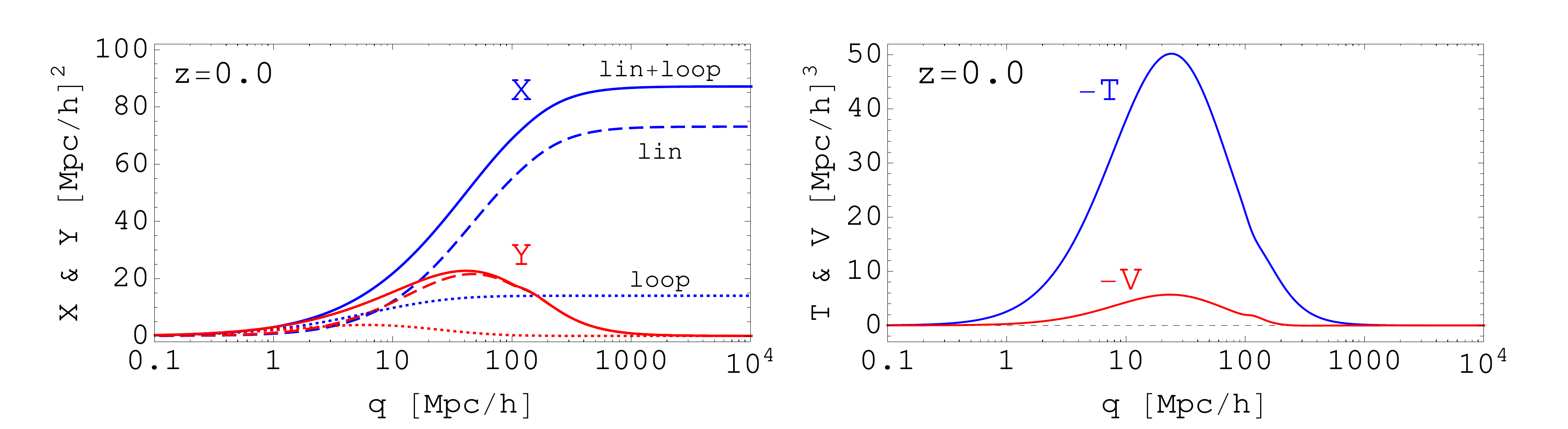}
   \end{center}
   \vspace*{-0.5cm}
   \caption{\small Scale dependence of two and tree point functions of the displacement field, equations 
   \eqref{eq:XYterm}, which contribute to the cumulant expansion, shown at redshift $z=0.0$. Linear (dashed) and 
   one loop (dotted) contributions to the $X$ (blue) and $Y$ (red) terms (solid line is linear + one loop) 
   are shown on the left panel. On the right panel we show tree level contribution to the  $V$ (red) and 
   $T$ (blue) terms (solid lines).}
   \label{fig:XYTV}
\end{figure*}

The cumulant expansion theorem allows expansion of the expected value of the exponential function
\begin{align}
K=\left\langle e^{-i\VEC{k}\cdot\Delta}\right\rangle
  =\exp\left[\sum^\infty_{N=1}\frac{(-i)^N}{N!}\left\langle 
  (\VEC{k}\cdot\Delta)^N\right\rangle_c\right],
\label{eq:cumexp}
\end{align}
where $\left\langle (\VEC{k}\cdot\Delta)^N\right\rangle_c$ stands for $N$-th cumulant of random 
variable. In diagrammatic representation this means that only connected terms contribute to the 
correlation. We can write
\cite{Carlson:2012bu}, 
\begin{align}
\log K&=\sum_{N=1}^\infty\frac{(-i)^N}{N!}\left\langle(\VEC{k}\cdot\Delta\right)^N\rangle_c\nonumber\\
&=\sum_{N=1}^\infty\frac{(-i)^N}{N!}k_{i_1}\ldots k_{i_N}\left\langle\Delta_{k_1}\ldots
\Delta_{k_N}\right\rangle_c.
\end{align}
In order to get the full power spectrum an infinite sum of these terms should be computed. 
However, since this is an expansion in powers of $k$, the series is convergent for sufficiently 
small values of $k$, in which case we can truncate the sum at a given order that meets our 
required accuracy. By isotropy we have that $N=1$ term vanishes. At the first order the 
displacement field is Gaussian and this gives the Zel'dovich approximation, for which only the 
$N=2$ cumulant is non-vanishing. In his paper we expand the displacement field to third order 
and keep only one loop terms, which means the fourth cumulant vanishes, hence we evaluate 
the summation to the third cumulant of $\Delta$ field, which leaves two cumulants to evaluate
\begin{align}
A_{ij}(\VEC{q})&=\left\langle\Delta_i\Delta_j\right\rangle_c,\nonumber\\
W_{ijk}(\VEC{q})&=\left\langle\Delta_i\Delta_j\Delta_k\right\rangle_c.
\label{eq:AWdef}
\end{align} 
This gives
\begin{align}
\log K=-\frac{1}{2}k_ik_jA_{ij}(\VEC{q})+\frac{i}{6}k_ik_jk_lW_{ijl}(\VEC{q}).\nonumber
\end{align}
Using this we have for the expression for the power spectrum given up to the third cumulant
\begin{align}
(2\pi)^3&\df^D (k)+P(k)=\int d^3 q ~e^{-i\VEC{q}\cdot\VEC{k}}\nonumber\\
&\times\exp\left[-\frac{1}{2}k_ik_jA_{ij}(\VEC{q})+\frac{i}{6}k_ik_jk_lW_{ijl}(\VEC{q})\right].
\label{eq:PSwithAW}
\end{align}
Next step is to evaluate the contributing cumulants in CLPT. 
The displacement cumulants $A_{ij}(\VEC{q})$ 
and $W_{ijk}(\VEC{q})$ can be decomposed into irreducible components relative to the pair 
separation vector $\VEC{q}$:
\begin{align}
A_{ij}(\VEC{q})&=X(q)\df^K_{ij}+Y(q)\hat{q}_i\hat{q}_j,\nonumber\\
W_{ijk}(\VEC{q})&=V(q)\hat{q}_{\left\lbrace i\right.}\df^K_{\left. jk\right\rbrace}+T(q)\hat{q}_i\hat{q}_j\hat{q}_k,
\label{eq:XYterm}
\end{align}
here, we have introduced the four scalar functions $X(q)$, $Y(q)$, $V(q)$ and $T(q)$ which depend 
on the amplitude of separation $q$. Angular brackets on the summation indexes imply that the 
summation is to be taken over all of the cyclic permutations. This follows from the fact that $W_{ijk}$ 
cumulant is symmetric under a permutation of its indexes. Contracting indexes on these tensors and 
solving the system we get
\begin{align}
\left.
\begin{array}{ll}
A_0\equiv\df^K_{ij}A_{ij}=3X+Y\\
\bar{A}\equiv\hat{q}_i\hat{q}_jA_{ij}=X+Y
\end{array} \right\rbrace
\rightarrow
\begin{array}{rl}
X&=\frac{1}{2}(A_0-\bar{A})\\
Y&=\frac{1}{2}(3\bar{A}-A_0),
\end{array}
\label{eq:XYdef} 
\end{align}
for the second cumulant, and similarly
\begin{align}
\left.
\begin{array}{ll}
W_0\equiv\hat{q}_i\df^K_{jk}W_{ijk}=5V+T\\
\overline{W}\equiv\hat{q}_i\hat{q}_j\hat{q}_kW_{ijk}=3V+T
\end{array} \right\rbrace
\rightarrow
\begin{array}{rl}
V&=\frac{1}{2}(W_0-\overline{W})\\
T&=\frac{1}{2}(5\overline{W}-3W_0).
\end{array} 
\label{eq:TVdef}
\end{align}
for the third cumulant. Using this we can rewrite the power spectrum into the form
\begin{align}
(2\pi)^3\df^D(k)&+P(k)=\int d^3 q ~e^{i\mu k\left(q-\frac{1}{2}k^2V\right)}\nonumber\\
&\times\exp\left[-\frac{1}{2}k^2(X+\mu^2 Y)-\frac{i}{6}\mu^3 k^3 T\right],
\label{eq:PSwithVT}
\end{align}
where we have introduced the angle between the given $k$-mode and separation vector 
$\mu=\hat{q}\cdot\hat{k}$.

It is worth keeping in mind that $A_{ij}$ is the two point correlator of the difference of displacement 
field and so contains a zero lag component: one can write
\begin{align}
A_{ij}(\VEC{q})&=\left\langle\Delta_i\Delta_j\right\rangle_c\nonumber\\
&=2\left(\sigma^2\delta_{ij}^K -\left\langle\Psi_i (\vq_1) \Psi_j (\vq_2) \right\rangle_{\vq_2-\vq_1 =\vq} \right), 
\end{align}
where $\sigma^2\delta_{ij}^K=\left\langle \Psi_i (\VEC{q})\Psi_j(\VEC{q}) \right\rangle=\frac{1}{2}
X(q \rightarrow \infty)$ is the squared zero lag rms displacement, i.e. displacement dispersion.
Because it is a zero lag quantity it is susceptible to nonlinear effects down to very small scales, where perturbation theory is unlikely to 
be reliable. Since this quantity does not depend on $\VEC{q}$ its Fourier transform is zero except for 
$k=0$. Because of this we will see below that it does not enter the final density power spectrum at 
the lowest order in $A_{ij}$, but it does enter at the quadratic order in $A_{ij}$ even in the low $k$ 
limit. In fact, due to its large value it dominates the nonlinear effects in this limit and is responsible 
for the smoothing of the BAO, among other effects. We will return to this discussion below.

\subsection{Perturbation theory of the displacement fields}
\label{subsec:ptfields}

We use Lagrangian perturbation theory (LPT) up to one loop to compute the contributions to scalar 
functions $X$, $Y$, $V$ and $T$. This has in most parts been derived in \cite{Carlson:2012bu} and 
we summarize it here for completeness, and in order to set up a framework for the later section 
\ref{sec:2LPTcross}, when we look at the cross power spectra of 2LPT. Detailed derivation of the 
$X$, $Y$ and $V$ $T$ terms is also given in appendix \ref{sec:XY}, and \ref{sec:VT}, respectively.

We start from the ansatz for the displacement field in Fourier space (see e.g. \cite{Matsubara:2007wj} )
\begin{align}
\Psi_i(\vp,\tau)&=\sum_{n=1}^\infty \Psi_i^{(n)}(\vp,\tau)\nonumber\\
&=-i\sum_{n=1}^\infty \frac{D^{(n)}(\tau)}{n!}
\int\prod_{l=1}^n\bigg[\frac{d^3p_l}{(2\pi)^3}\df_L(\vp_l)\bigg]\nonumber\\
&\times(2\pi)^3\df^3\Big(\sum_{j=1}^n\vp_j-\vp\Big)L^{(n)}_i(\vp_1,\ldots,\vp_n).
\label{eq:psiansatz}
\end{align} 
where $\df_L$ is the linear dark matter density field. Plugging this ansatz into the equation of motion 
and consistently solving order by order one gets the solution for the vector displacement kernels 
$\VEC{L}^{(n)}(\vp_l)$. This gives (see e.g. \cite{Catelan:1994ze, Catelan:1996hw, Bernardeau:2001qr, 
Matsubara:2007wj, Rampf:2012xa} 
\begin{align}
&L^{(1)}_i=\frac{k_i}{k^2},\nonumber
\end{align}
\begin{align}
&L^{(2)}_i\left(\vp_1,\vp_2\right)=
\frac{3}{7}\frac{k_i}{k^2}\bigg[1-\Big(\frac{\vp_1\cdot\vp_2}{p_1p_2}\Big)^2\bigg],\nonumber\\
&L^{(3)}_i\left(\vp_1,\vp_2,\vp_3\right)=
\frac{5}{7}\frac{k_i}{k^2}\Bigg[1-\bigg(\frac{\vp_2\cdot\vp_3}{p_2p_3}\bigg)^2\Bigg]\nonumber\\
&~~~\times\Bigg\lbrace 1-\bigg[\frac{\vp_1\cdot (\vp_2+\vp_3) } {p_1|\vp_1+\vp_2|}\bigg]^2\Bigg\rbrace
-\frac{1}{3}\frac{k_i}{k^2}\Bigg[1-3\bigg(\frac{\vp_1\cdot\vp_2}{p_1p_2}\bigg)^2\nonumber\\
&~~~+2\frac{(\vp_1\cdot\vp_2)(\vp_2\cdot\vp_3)(\vp_3\cdot\vp_1)}{p_1^2p_2^2p_3^2}\Bigg]
+\epsilon_{ijl}k_jK_l\left(\vp_1,\vp_2,\vp_3\right),
\end{align}
where $\vk=\vp_1+\ldots+\vp_n$ for $\VEC{L}^{(n)}$, and $K_l$ is the transverse part which does 
not enter at the lowest order. For the last $\VEC{L}^{(3)}$ kernel it is useful to make it fully symmetrical 
in all the $\vp_i$ variables. In general we can also solve for the time evolution of these kernels, i.e. solve the 
second order differential equation for each $D^{(n)}(\tau)$ (see e.g. \cite{Bernardeau:2001qr}), but for 
simplicity we assume the logarithmic growth rate to be $f(\tau)=d\ln D/d\ln a=\Omega_m^{1/2}(\tau)$. 
This simplifies the situation so the growth rate at each order in perturbations can be written as powers 
of linear growth rate $D^{(n)}(\tau)=D_L^{n}(\tau)$. 

As done in \cite{Matsubara:2007wj} it is useful to define multi-spectra of the displacement field
\begin{align}
&\left<\Psi_{i_1}(\VEC{p}_1)\ldots\Psi_{i_N}(\VEC{p}_1)\right>_c = (2\pi)^3\df^D(\VEC{p}_1+\ldots+
\VEC{p}_N) \nonumber\\
&~~~~\qquad\qquad\qquad\qquad\times i^{N-2}C_{i_1\ldots i_N}(\VEC{p}_1,\ldots,\VEC{p}_N),
\label{eq:npsi}
\end{align}
where $\Psi_i(\VEC{p})$ are the Fourier transforms of the displacement fields. 

Using this we can compute the $X$, $Y$ $V$ and $T$ terms up to one loop. Details of this calculation 
are presented in appendix \ref{sec:XY} and \ref{sec:VT}, and can also be found in e.g. \cite{Carlson:2012bu}. 
In figure \ref{fig:XYTV} we show the result of up to one loop prediction of these terms at redshift $z=0.0$. 
We see that going beyond the Zel'dovich calculation introduces the corrections to the $X$ and $Y$ terms 
where for $Y$ term we see that corrections are restricted to the scales below $\sim100~\text{Mpc}/h$, while 
for the $X$ term on the other hand we have a correction on very large scales, which means that the one loop 
calculation gives a considerable contribution to the zero lag rms displacement. $V$ and $T$ terms are pure 
one loop terms which are zero in the linear approximation. Both terms asymptote to zero at large and small scales. 
For $Y$, $V$, and $T$ terms we see that they have a peak at the scales of around $\sim30~\text{Mpc}/h$.

\subsection{Expansion in the angular moments}
\label{subsec:angmom}

\begin{figure*}[t]
   \begin{center}
   \hspace*{-0.5cm}
   \includegraphics[scale=0.47]{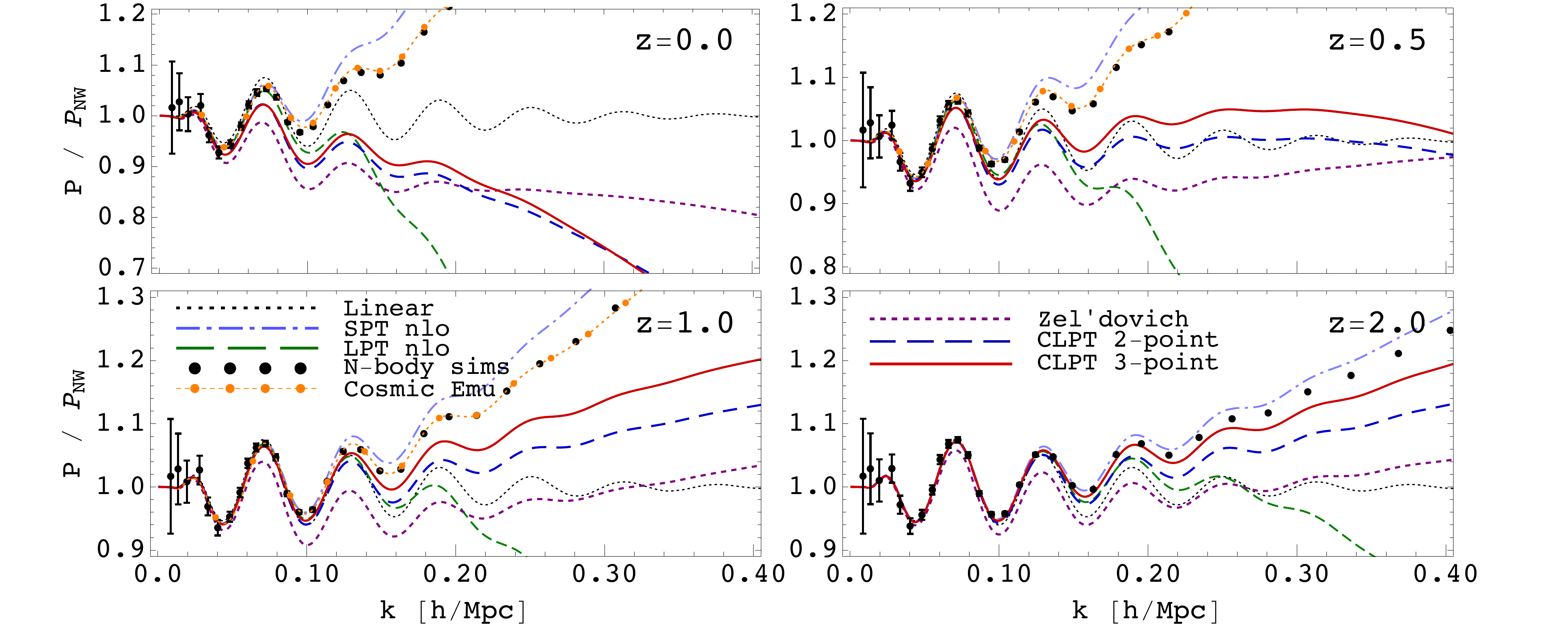}
   \end{center}
   \vspace*{-0.5cm}
   \caption{\small Power spectrum result obtained by several methods at redshift $z=0.0,~0.5,~1.0$ and $2.0$.
   Full CLPT result of equation \eqref{eq:PSwithVT} at one loop is shown (solid red line) together with the approximations 
   where tree point contribution of $V$ and $T$ terms are dropped and two point terms $X$ and $Y$ remain at 
   one loop (long-dashed blue line). Shown is also the corresponding result presented in \cite{Carlson:2012bu} 
   (long-dashed red line), where the exponent with three point term $W_{ijk}$ is expanded and only the first term 
   is kept. This turns out to be a good approximation on scales shown here, and the difference is hardly noticeable. 
   We also show the Zel'dovich result (short-dashed purple line), usual one loop 
   SPT (dot-dashed light-blue line), one loop LPT (dashed green line) as presented in \cite{Matsubara:2007wj} as 
   well as linear theory (dotted black line). For comparison we show the N-body simulation results (black dots) and 
   cosmic emulator results \cite{Lawrence:2009uk} (orange connected-dots). All the spectra are divided by the 
   no-wiggle linear power spectra \cite{Eisenstein:1997ik} in order to reduce the range of scales.}
   \label{fig:PowerSpectrum}
\end{figure*} 

It is known that evaluating the matter power spectrum even in the Zel'dovich approximation is not as straightforward 
as doing the direct two dimensional integral. Direct numerical integration is not the optimal approach since the 
integral function can be highly oscillatory. In \cite{Schneider:1995} the method is proposed to solve the Zel'dovich 
approximation power spectrum. Here  we generalise this method to evaluate the higher order power spectrum. 

We can express the power spectrum in the spherical frame where $\VEC{k}$ is along the $\hat{z}$ direction:
\begin{align}
 P(k)&=2\pi\int_0^\infty q^2dq~\int_{-1}^{1} d\mu~e^{i\mu k q}\left\lbrace e^{-\frac{1}{2}k^2X} 
 e^{-\frac{i}{2}\mu k^3 V} \right.\nonumber\\
&\left. \times\exp\left[-\frac{1}{2}\mu^2k^2Y-\frac{i}{6}\mu^3 k^3 T\right]-e^{-\frac{1}{2}k^2\sigma^2}\right\rbrace,
\label{eq:explicitePS}
\end{align}
where in the last term we added the zero lag term, which is an extra contribution that is a constant, hence 
vanishes for all $k$ except $\VEC{k}=0$. It is introduced to cure the oscillatory integration problems. It depends on 
$\sigma^2$, the squared rms displacement, which can be evaluated in high $q$ limit of $X(q)$ (figure \ref{fig:XYTV}).
We will return to this in the next section where we will focus on the low $k$ limit of the power spectrum. 
Direct evaluation of equation \eqref{eq:explicitePS} is difficult because of fast oscillating terms. Instead we first
rewrite equation \eqref{eq:PSwithVT} in more convenient form for evaluation
\begin{align}
P(k)=\mathcal{Z}(k)+\mathcal{V}(k)+\mathcal{T}(k)
\end{align}
where we have
\begin{align}
\mathcal{Z}(k)&=\int d^3 q ~e^{i\mu k q}\left(e^{-\frac{1}{2}k^2(X+\mu^2 Y)}-e^{-\frac{1}{2}k^2\sigma^2}\right),\nonumber\\
\mathcal{V}(k)&=\int d^3 q ~e^{i\mu k q}e^{-\frac{1}{2}k^2(X+\mu^2 Y)}\left(e^{-\frac{i}{2}\mu k^3V}-1\right),\nonumber\\
\mathcal{T}(k)&=\int d^3 q ~e^{i\mu k\left(q-\frac{1}{2}k^2V\right)} \nonumber\\
&\qquad ~~~ \times e^{-\frac{1}{2}k^2(X+\mu^2 Y)}\left(e^{-\frac{i}{6}\mu^3 k^3 T}-1\right)
\label{eq:ZVT}
\end{align}
Note that the first contribution $\mathcal{Z}$ is the nonlinear Zel'dovich case where only the two point 
contribution in the cumulant expansion is considered. To evaluate these terms we can use the expansion 
formula presented in Appendix \ref{sec:int1}. For the first two terms above, $\mathcal{Z}$ and $\mathcal{V}$ 
we can use the expansion
\begin{align}
\int^1_{-1} d\mu~e^{iA\mu}~\exp(B\mu^2)=2e^B \sum_{n=0}^\infty \left(-\frac{2B}{A}\right)^nj_n(A),
\label{eq:intAB}
\end{align}
and for the third term we use the generalized equation \eqref{eq:intAB} 
\begin{align}
\int^1_{-1} d\mu~ e^{iA\mu}~\exp(B&\mu^2 + i\epsilon\mu^3)=\nonumber\\
&2 e^B \sum_{n=0}^\infty \left(-\frac{2B}{A}\right)^n J_n(A,\epsilon),
\label{eq:intABe}
\end{align}
where we $J_n(A,\epsilon)$ is the generalization for the spherical Bessel function $j_n(A)$ which we had in
previous case. Explicit form for $J_n$ is given by equation \eqref{eq:newJ}. Note that in the limit of $\epsilon 
\rightarrow 0$ we retrieve the result above, i.e. $J_n(A,\epsilon) \rightarrow j_n(A)$.
We see that integrals in equation \eqref{eq:ZVT} can be expressed in terms of these expansions using 
\begin{align}
A(k,q)&=k\left(q-\frac{1}{2}k^2V(q)\right),\nonumber\\
B(k,q)&=-\frac{1}{2}k^2 Y(q),\nonumber\\
\epsilon (k,q)&= -\frac{1}{6}k^3 T(q).
\end{align}
Doing so we have reduced the equation \eqref{eq:ZVT} integrals from three dimensional integrals 
to a quickly converging sum of one dimensional integrals. Typically the sum over $n$ can be 
truncated at $n<15$ for $k<1 h/\text{Mpc}$ (in \cite{Schneider:1995} it was argued $n=3$ is good 
enough for $k<0.3 h/\text{Mpc}$). Since one dimensional integration over $q$ for a given $k$ is fast 
we use a conservative value of $n=25$. We also developed an alternative expansion in spherical 
harmonics, which is presented in Appendix \ref{sec:int2}. This gives equivalent numerical results 
and will not be discussed here in more detail.

In figure \ref{fig:PowerSpectrum} we show the results of the one loop CLPT power spectrum for 
four different redshifts, $z=0.0$, $0.5$, $1.0$ and $2.0$. Also shown are the results when three 
point function $W_{ijk}$ in equation \eqref{eq:AWdef} is neglected and only $A_{ij}$ term remains. 
We compare these to the N-body results as well as one loop SPT results. We see that CLPT at low redshifts 
is significantly below the N-body results. We also investigate the corresponding result presented in 
\cite{Carlson:2012bu}, where the exponent with three point term $W_{ijk}$ is expanded and only 
the first term is kept. On the mildly-nonlinear scales that we are showing this is a good approximation 
and it can hardly be distinguished from the full result presented in the same figure \ref{fig:PowerSpectrum}.
Comparison of this linear approximation to the full result is shown in the appendix \ref{sec:int2} and in figure \ref{fig:VT}. 
Adding the three point function helps in the sense that it adds power, but the effect is relatively small. 
We also see that the effect of adding one loop corrections to Zeldovich leads to an increase in power at higher 
redshifts and also at lower redshifts for low $k$, as desired, but actually reduces power at higher $k$ for 
lower redshifts. This can be interpreted as a sign of things gone astray in this approach. We will address this issue 
again below, pointing out that the zero lag rms displacement correction in one loop calculations yields too large contribution. 

\subsection{Low $k$ limit}
\label{sec:lowk}

In this section we expand our result in equation \eqref{eq:PSwithVT} in $k$
powers to get the $k^2$ corrections to the linear theory. 
Expanding equation \eqref{eq:PSwithAW} it follows
\begin{align}
(2\pi)^3\df^D(k)&+P(k)
=(2\pi)^3\df^D(k) - \frac{1}{2}k_ik_j\int d^3 q ~e^{i\VEC{q}\cdot\VEC{k}}A_{ij}\nonumber\\ 
&-\frac{i}{6} k_ik_jk_l\int d^3 q ~e^{i\VEC{q}\cdot\VEC{k}}W_{ijl}\nonumber\\ 
&+\frac{1}{8} k_ik_jk_lk_m \int d^3 q ~e^{i\VEC{q}\cdot\VEC{k}}A_{ij}A_{lm}+\ldots
\label{eq:Plimit}
\end{align}
Evaluating each of these terms using equations \eqref{eq:XYex}, \eqref{eq:VTex} and the 
standard identities for spherical Bessel functions (e.g. \cite{NIST:DLMF,Olver:2010:NHMF}) gives
\begin{align}
- \frac{1}{2}k_ik_j\int d^3 q ~e^{i\VEC{q}\cdot\VEC{k}}A_{ij}=&p(k)\\
-\frac{i}{6} k_ik_jk_l\int d^3 q ~e^{i\VEC{q}\cdot\VEC{k}}W_{ijl}=&\frac{3}{7}\Big(Q_2(k)+2R_2(k)\Big)\nonumber\\
\frac{1}{8} k_ik_jk_lk_m \int d^3 q ~e^{i\VEC{q}\cdot\VEC{k}}A_{ij}A_{lm}=&\frac{1}{2}Q_3(k)-k^2\sigma^2 p(k),\nonumber
\end{align}
where we use
\begin{align}
p(k)=P_L(k)+\frac{9}{98}Q_1(k)+\frac{10}{21}R_1(k).
\label{eq:malip}
\end{align}
Here all $Q_i$ and $R_i$ are as defined in \cite{Matsubara:2007wj}. For example, we have
\begin{align}
\bar{Q}_3(k)=&\int{\frac{d^3 k'}{(2\pi)^3}
\frac{(\VEC{k}\cdot\VEC{k}')^2}{k'^4}\frac{(\VEC{k}\cdot(\VEC{k}-\VEC{k}'))^2}{(\VEC{k}-\VEC{k}')^4}}\nonumber\\
&\qquad\qquad\qquad\times p(k')p(|\VEC{k}-\VEC{k}'|).
\end{align}

\begin{figure*}[t!]
   \begin{center}
   \hspace*{-0.5cm}
   \includegraphics[scale=0.505]{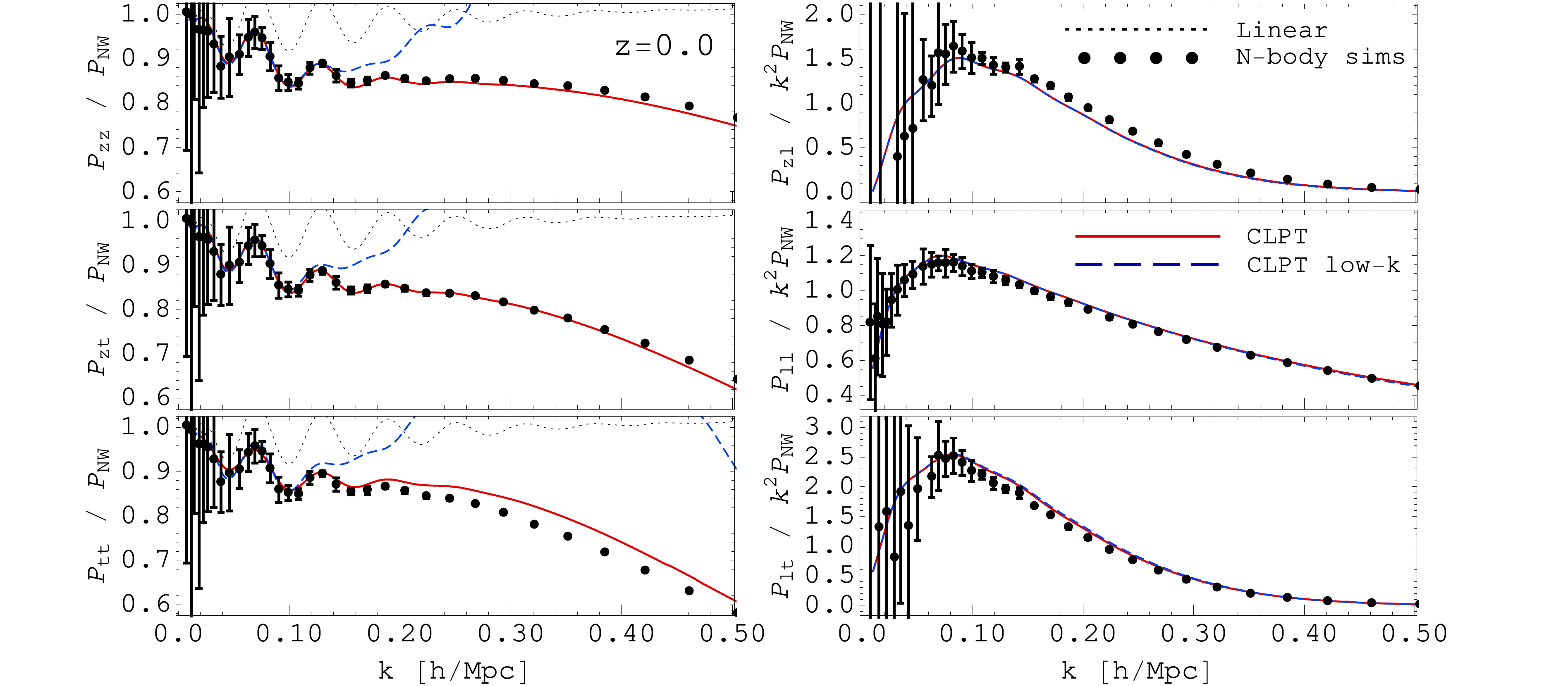}
   \end{center}
   \vspace*{-0.5cm}
   \caption{\small Cross and auto power spectrum results from table \ref{tb:crossautoPS} using up to the 2LPT 
   displacement at redshift $z=0.0$. Result of equation \eqref{eq:PSwithVT} up to the 2LPT is shown (solid red lines) 
   together with the low-$k$ limit results (dashed blue lines). For comparison we show the measured 2LPT simulation power 
   spectrum (black dots) obtained by displacing particle on the grid with the initial condition codes \cite{Crocce:2006ve}.
   All the spectra are divided by the no-wiggle linear power spectra \cite{Eisenstein:1997ik} in order to reduce the range 
   in the plots. 
   }
   \label{fig:CrossPS}
\end{figure*} 

It is useful to define the 
projector operator $\Pp$ which acts on a function by projecting the full one loop 
result to its perturbation components: linear part  $\Pp_L$, convolution part $\Pp_Q$
and the propagator part $\Pp_R$. As an example, the result of applying 
the $\Pp_Q$ to the full one loop result of bispectrum term $T$ given in equation 
\eqref{eq:VTex} is
\begin{align}
\mathbb{P}_Q T(q) = 3\int \frac{dk}{2{\pi}^2 k}\left(-\frac{3}{7}\right)\Big(Q_1(k)+2Q_2(k)\Big)j_3(qk),
\end{align}
Note that summing the three different operators gives identity operation, i.e. $\Pp_L+\Pp_Q+\Pp_R=\mathbb{I}$.
With this one has
\begin{align}
\sigma^2=\sigma^2_L+\sigma^2_\text{1loop}=\frac{1}{3}\int{\frac{d^3k}{(2\pi)^3}\frac{p(k)}{k^2}}.
\end{align}
Here we have used the labels for the dispersion contributions $\sigma^2_L=\Pp_L\sigma^2$ and
$\sigma^2_\text{1loop}=(\Pp_Q+\Pp_R)\sigma^2$. If we also define the new contributions
\begin{align}
Q_{3Q,L}(k)=&\int{\frac{d^3 k'}{(2\pi)^3}
\frac{(\VEC{k}\cdot\VEC{k}')^2}{k'^4}\frac{(\VEC{k}\cdot(\VEC{k}-\VEC{k}'))^2}{(\VEC{k}-\VEC{k}')^4}}\nonumber\\
&\qquad\qquad\qquad\times\Pp_{Q,L} p(k')\Pp_{Q,L} p(|\VEC{k}-\VEC{k}'|)
\label{eq:sig}
\end{align}
This gives the correction to the one loop SPT power spectrum
\begin{align}
P(k)=P^{\text{SPT}}_{\text{1loop}}(k)+\frac{1}{2}\left(Q_3(k)-Q_{3L}(k)\right)-k^2P_L(k)\sigma^2_\text{1loop}.
\label{eq:spt}
\end{align}
The last part here gives the correction to the $k^2$ SPT propagator which suppresses the power. 
In the high $k$ limit this term is cancelled by the second term due to the fact that 
relative displacement field vanishes in the limit of a small separation (the so called Galilean  
invariance), but in the low $k$ limit the last term dominates. 
At $z=0$ the linear theory value of $\sigma_{L}^2=36.55~(\text{Mpc}/h)^2$, 
and $\sigma_\text{1loop}^2=7.00~(\text{Mpc}/h)^2$, so the one loop correction is quite large. 

In this paper we argue that zero lag quantities are difficult to evaluate perturbatively because they 
receive contributions from all scales, including very small scales not amenable to the perturbation theory. 
Both the linear Zel'dovich and its one loop generalization suffer from the adhesion problem: while in simulations 
particles stop its displacement streaming because they are trapped inside the dark matter halos, in 
Zeldovich approximation and its higher order LPT extensions this does not happen and the particles keep 
streaming along their paths. Because of this the displacement field will be filtered out on small scales, 
which is the regime where one loop calculation predicts a large contribution. 
As a consequence, one loop LPT is unreliable in its rms displacement prediction: results from N-body simulations 
(see e.g. \cite{Chan:2013vao, Tassev:2013rta} suggest that the full nonlinear value should be comparable or
slightly higher 
than the linear prediction $\sigma_L^2$. This is because most of the linear Zel'dovich contribution comes from 
rather large scales, where the predictions are reliable. If we erase both of the last two terms in equation \eqref{eq:spt}
we obtain precisely the one loop SPT result in this limit. If, instead, we neglect the second term and allow for a free 
$\sigma^2$ in the last term of equation \eqref{eq:spt}, i.e. if we replace $\sigma^2$ with a free parameter, we retrieve 
the same correction form to the SPT as in one loop effective field theory (e.g. \cite{Carrasco:2013mua}). 

\begin{table*}[tp!]
\caption{Cross and auto power spectra up to 2LPT.} 
\centering 
\setlength{\tabcolsep}{8pt}
\renewcommand{\arraystretch}{1.0}
\begin{tabular}{c|cccccc}
\hline\hline
  & $P^{zz}$ & $P^{zl}$ & $P^{zt}$ & $P^{ll}$ & $P^{lt}$ & $P^{tt}$\\ [0.5ex] 
\hline
$X^{\alpha\beta}$ & $\Pp_L X$ & $(\Pp_L+\Pp_Q)\sigma^2$ & $\Pp_L X+\Pp_Q\sigma^2$ 
& $\Pp_Q X$ & $\Pp_Q X+\Pp_L\sigma^2$ & $(\Pp_L+\Pp_Q) X$\\
$Y^{\alpha\beta}$ & $\Pp_L Y$ & $0$ & $\Pp_L Y$ & $\Pp_Q Y$ & $\Pp_Q Y$ &  $(\Pp_L+\Pp_Q) Y$\\
$V^{\alpha\beta}$ & $0$ & $\frac{1}{2}\mathbb{P}_QV$ & $\frac{1}{2}V$ & $0$ & $\frac{1}{2}\mathbb{P}_QV$ & $V$\\
$T^{\alpha\beta}$ & $0$ & $\frac{1}{2}\mathbb{P}_QT$ & $\frac{1}{2}T$ & $0$ & $\frac{1}{2}\mathbb{P}_QT$ & $T$\\
\hline
\end{tabular}
\label{tb:crossautoPS}
\end{table*}

\subsection{Cross and auto power at 2LPT}
\label{sec:2LPTcross}

Natural question that emerges by looking at the equation \eqref{eq:PSwithAW} is related to how good is the truncation
of the cumulant expansion, and how well does equation \eqref{eq:PSwithAW} preform assuming perfectly modeled
$X$, $Y$, $V$ and $T$ terms. In order to answer that one would need an accurate simulation measurements of 
all these terms. Alternative is to perform a similar test on different (simpler) $X$, $Y$, $V$ and $T$ terms for which 
the solution of equation \eqref{eq:PSwithAW} can be obtained, and final result can be cross checked with direct numerical 
calculation of the power spectrum on the grid. In this scenario we assume that chosen terms $X$, $Y$, $V$ and $T$ do not 
differ from the previous (nonlinear evolution) case in any pathological way, which might alter the final conclusion.

In order to perform this test we can use LPT displacements to compare the performance of the analytical solution of 
the integral in equation \eqref{eq:PSwithAW}, to the cross and auto power spectra obtained from `initial condition' code 
\cite{Crocce:2006ve} at redshift $z=0$, where the nonlinear effects are most apparent. For this purpose let us first introduce 
labelling that we will use in analysing all of the cross and auto spectra. Let us identify the set of indexes $\left\lbrace z,l,t\right\rbrace=
\left\lbrace \text{zel},\text{2lpt},\text{zel}+\text{2lpt}\right\rbrace$, where in the first case we use the Zel'dovich result, 
followed by the sole 2LPT result and finally a sum of the two. Using the same formalism as when deriving equation 
\eqref{eq:PSwithAW}, we can write 
\begin{align}
(2\pi)^3\df^D (k)&+P^{\alpha\beta}(k)=\int d^3 q ~e^{-i\VEC{q}\cdot\VEC{k}}\\
&\times\exp\left[-\frac{1}{2}k_ik_jA^{\alpha\beta}_{ij}(\VEC{q})+\frac{i}{6}k_ik_jk_lW^{\alpha\beta}_{ijl}(\VEC{q})\right]\nonumber.
\end{align}
Both, $\alpha$ and $\beta$, indexes can take the values from the set $\left\lbrace z,l,t\right\rbrace$ in 
representing the Zel'dovich, 2LPT or combined result. Note that the $P^{zz}$ result is the standard
Zel'dovich power spectrum, using the linear displacement.  In table \ref{tb:crossautoPS} we show the 
result for all the combinations of cross and auto spectra up to the 2LPT. We show the results in terms 
of how the decomposition coefficients of $X$, $Y$, $V$ and $T$ change when using different 
combination of PT approximation levels. Using equations \eqref{eq:Plimit} and result of the table 
\ref{tb:crossautoPS} we also can find the low $k$ limit for each of the cross power spectra. We have
\begin{align}
P^{zz}&=(1-k^2\sigma_{L}^2)P_L +\frac{1}{2}Q_3 ,\nonumber\\
P^{zl}&=e^{-\frac{1}{2}k^2(\sigma_{L}^2+\sigma_{Q}^2)}\frac{3}{14}Q_2,\nonumber\\
P^{zt}&=e^{-\frac{1}{2}k^2\sigma_{Q}^2}\Big((1-k^2\sigma_L^2)P_L+\frac{3}{14}(Q_2+2R_2)+\frac{1}{2}Q_3\Big),\nonumber\\
P^{ll} &=(1-k^2\sigma_Q^2)\frac{9}{98}Q_1+\frac{1}{2}Q_{3Q},\nonumber\\
P^{lt}&=e^{-\frac{1}{2}k^2\sigma_{L}^2}\Big((1-k^2\sigma_Q^2)\frac{9}{98}Q_1+\frac{3}{14}Q_2+\frac{1}{2}Q_{3Q}\Big),\nonumber\\
P^{tt}&=\left(1-k^2(\sigma_L^2+\sigma_Q^2)\right)\left(P_L+\frac{9}{98}Q_1\right)\nonumber\\
&\qquad\qquad\qquad\qquad+\frac{3}{7}(Q_2+2R_2)+\frac{1}{2}Q_3.
\label{eq:lowkP}
\end{align}

\begin{figure*}[t!]
   \begin{center}
   \hspace*{-0.5cm}
   \includegraphics[scale=0.47]{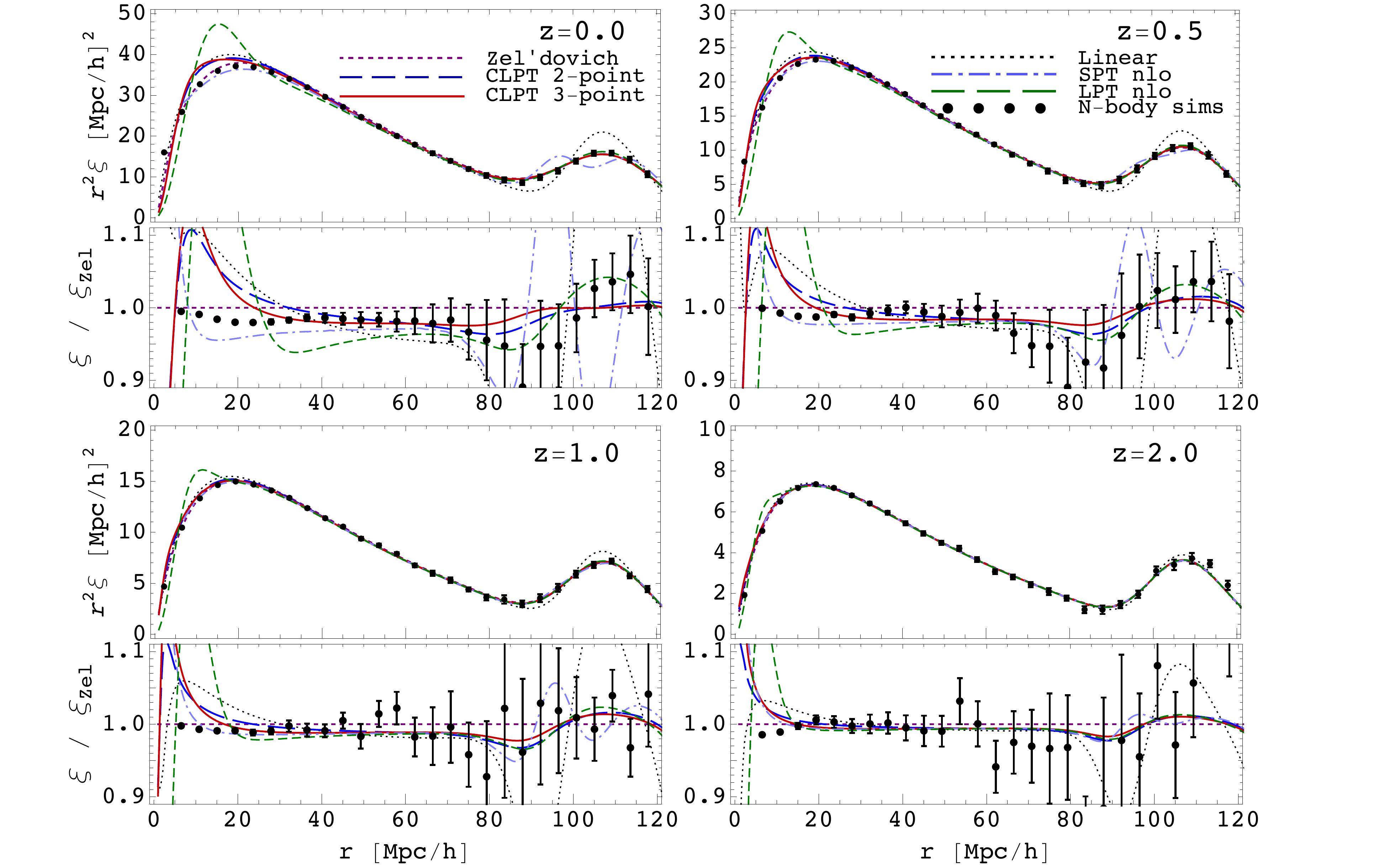}
   \end{center}
   \vspace*{-0.5cm}
   \caption{\small Correlation function obtained by numerically Fourier transforming the results from figure \ref{fig:PowerSpectrum}
   at redshift $z=0.0,~0.5,~1.0$ and $2.0$. In the upper panel we show the correlation function and in the lower panel the same 
   result is divided by the Zel'dovich result (short-dashed purple line in upper panel). 
   Full result of equation \eqref{eq:PSwithVT} at one loop is shown (solid red line) together with the approximations where tree point 
   contribution of $V$ and $T$ terms are dropped and two point terms $X$ and $Y$ remain at one loop (long-dashed blue 
   line). For comparison we also show one loop LPT (short-dashed green line) as presented in \cite{Matsubara:2007wj} as well as linear 
   theory (dotted black line). For comparison we also show the N-body simulation results (black dots).
   }
   \label{fig:CorrFunction}
\end{figure*} 

In figure \ref{fig:CrossPS}  we show the results of 6 different cross and auto spectra up to the 2LPT
approximations. We compare these results with numerical results obtained from measuring the 
power spectra on the grid from initial condition code \cite{Crocce:2006ve}. For N-body grid
results we use $1000$ Mpc$/h$ box size with $1024^3$ particles. Since we are dealing with the finite box size
there are some residual effects. For example there will be effects coming from Nyquist frequency cut off in N-body result
which we do not account for in analytical calculations. Up to these numerical effects 
we see that overall, for all of the cross spectra we are considering, the analytical 
calculations and numerical measurements agree well with each other. In addition to the numerical effects 
mentioned above in some of the spectra like $P_{ll}$, $P_{lt}$ and $P_{tt}$ the contribution of 2LPT bispectrum 
term $\left\langle\Psi^{(2)}\Psi^{(2)}\Psi^{(2)}\right\rangle$ can be considered. Since this is formally a two 
loop term it does not enter into our analytic result but it is present in the numerical N-body  result. 

From the results in figure \ref{fig:CrossPS}  we conclude that equation \eqref{eq:PSwithAW} agrees well with 
the N-body grid solution, given the same $X$, $Y$, $V$ and $T$ terms. Differences observed between these 
two solutions are significantly smaller than what we have seen in the fully nonlinear when compared to
full N-body simulation in figure \ref{fig:PowerSpectrum}.  This leads the conclusion that in order to improve
the modeling of the analytic solution we need to turn to more accurate modeling of the displacement field 
power spectra that contribute to $X$, $Y$, $V$ and $T$ terms. 

\subsection{Correlation function}
\label{sec:corrfnc}

Any well defined PT model should allow for a comparison of results both in Fourier space and in configuration space. 
Correlation function is defined as the two point correlation of density field in configuration 
space and it is a Fourier transform of the power spectrum
\begin{align}
\xi(r)&=\int{\frac{d^3k}{(2\pi)^3} ~\text{exp}(-i\VEC{k}\cdot\VEC{r})P(k)}\nonumber\\
&=\int{\frac{k^2d k}{2\pi^2}~P(k)j_0(kr)}
\end{align} 
We Fourier transform the main results shown in figure \ref{fig:PowerSpectrum}.
In figure \ref{fig:CorrFunction} we show the results for the correlation 
function at redshifts $z=0.0$, $z=0.5$, $z=1.0$ and $z=2.0$.  We show the results of the 
equation \eqref{eq:PSwithVT} at one loop together with the approximations where tree point 
contribution of $V$ and $T$ terms are dropped  and two point terms $X$ and $Y$ remain 
at one loop. For comparison we also show one loop LPT  from \cite{Matsubara:2007wj}, where 
PT terms are not kept in the exponential but expanded, and which 
is preforming worse than both one loop calculation CLPT and Zel'dovich. 
As mentioned before, formally the result in this paper differs from the one presented 
in \cite{Carlson:2012bu} in respect that there the exponent with three point function 
is expanded and only the leading term (linear in $V$ and $T$) is kept, and in this work we 
keep all the terms in the exponent. In practice though these two methods give very similar 
results since the corrections coming from the expansion terms above the leading one are 
small on scales shown in the plots, and thus the results agree. 

\begin{figure*}[!t]
    \centering
    \includegraphics[trim = 0.43cm 0cm 0cm 0cm, clip, scale=0.59]{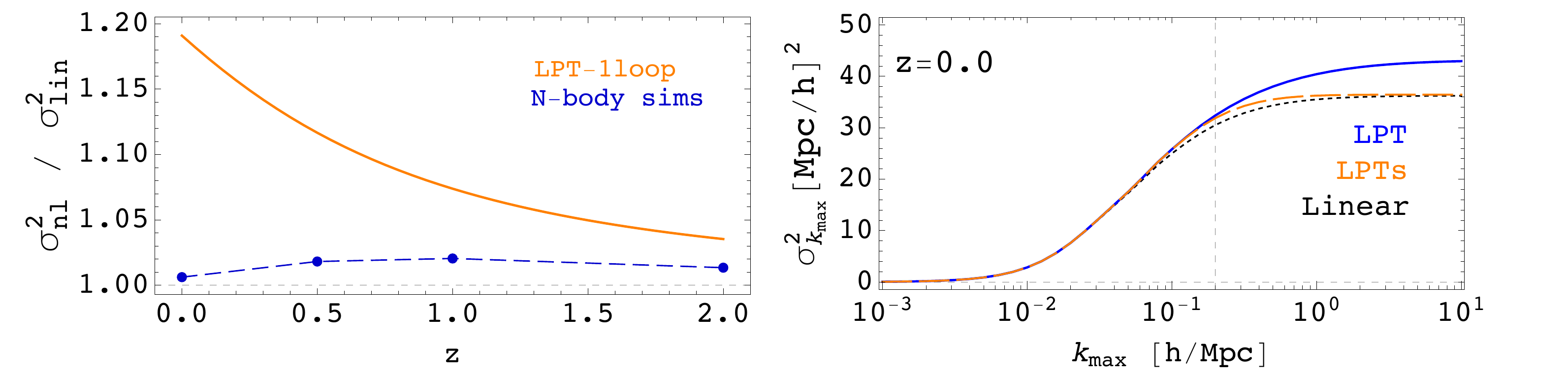}
    \caption{\small On the \textit{left panel} we show the ratio of nonlinear displacement dispersion $\sigma^2$ 
    and linear theory prediction, as a function of redshift $z$. We show the results measured in the N-body 
    simulations (blue dots and dashed line) as well as one loop LPT results (orange dots and dashed line).
    On the \textit{right panel} we show predictions of $\sigma^2_{k_{\text{max}}}$ at z=0.0, as a function of 
    the integration cutoff $k_{\text{max}}$, using linear theory (black dotted line), one loop LPT (blue solid line) and LPTs
    (orange dotted line).
    }
    \label{fig:Sigmaplot}
\end{figure*}

From figure \ref{fig:CorrFunction} 
we see that one loop results improve the comparison against N-body simulations relative to 
Zel'dovich on scales around $40$ Mpc$/h$ and larger. The agreement around $40-60$Mpc/h is particularly impressive. 
Around BAO one loop CLPT does better than 
Zel'dovich, but seems to still be missing something. Below $40$ Mpc$/h$ one loop CLPT does considerably 
worse than Zel'dovich, even though in the power spectrum it showed an improvement at higher $k$ relative to 
Zel'dovich. We will address this further in the next section, where we present a model that improves upon 
our one loop calculations. 

\section{Beyond CLPT}
\label{sec:decomp}

The results presented in the previous section have shown that one loop CLPT is an improvement over 
Zel'dovich in the power spectrum, 
but one loop CLPT is still well below full power spectrum in simulations. 
On the other hand, correlation function results shown in figure \ref{fig:CorrFunction} paint a 
different picture, one where the Zel'dovich approximation gives a much better 
agreement with the N-body simulations than linear theory, and in some cases even better than 1 loop 
CLPT. This manifests itself 
particularly in the BAO smoothing, where little excess power remains in the N-body simulation 
data against Zel'dovich. 
The difference between the power spectrum and correlation function 
suggests that the power spectrum is strongly affected by very small scale correlations, which is 
difficult to get it right in CLPT. 
As an example, in the limit of a large contribution from 
small scale correlations limited to zero lag this becomes a shot noise term, which can completely change the power spectrum 
at all $k$, while not changing the correlation function at any non-zero lag value of $r$. 
The effects on the power spectrum can thus be very different from those on the correlation function. 

A second, and separate, issue is that realistic power spectrum
has narrow BAO features that also get modified by nonlinear evolution. To a large extent this is an 
easier problem in the context of CLPT since Zel'dovich approximation already reproduces nonlinear 
BAO effects quite well.  We will split this problem 
from the broadband problem, and decompose the total power spectrum into the Zel'dovich contribution,  
the residual BAO wiggle and residual broad band contributions. We can write  
\begin{align}
P(k)=P_{\text{Zel}}(k)+P_{W}(k)+P_{BB} (k),
\label{eq:PSmodel}
\end{align}
where the first term is the Zel'dovich power spectrum, second $P_{W}$ term is the BAO wiggle residual, 
and $P_{BB}$ is the residual broad band power. 

From the comparison of power spectrum results with N-body simulation results found in \cite{Chan:2013vao, 
Tassev:2013pn} we have learned that one loop LPT overestimates the displacement power spectrum at 
small scales (see figure 4 in \cite{Chan:2013vao}). As a consequence the total rms displacement field 
is also overestimated. In the left panel in figure \ref{fig:Sigmaplot} we show the result for rms displacement field 
measured in N-body simulations and compared to linear theory and one loop LPT prediction. We see that 
one loop LPT prediction overestimates the displacement field at all redshifts up to $z=2.0$. We note that 
for all redshifts simulations suggest that rms displacement is closer to linear theory than one loop LPT.
This indicates that result in equation \eqref{eq:spt} does not provide a correct low $k$ limit.
Physically this is a consequence of dark matter particles being trapped inside the dark matter 
halos, rather than streaming along their Zel'dovich or LPT trajectories. This effect is 
not captured at one loop LPT. On the right panel of figure \ref{fig:Sigmaplot} we show the contribution
to the rms displacement as a function of scale: 
\begin{align}
\sigma^2(k_{max}) = \int_0^{k_{max}}\frac{d \ln k}{6\pi^2}k~p(k),
\end{align}
where $p(k)$ is the displacement field power spectrum (for one loop LPT result see equation \eqref{eq:malip}). 
We can see that a significant contribution to the one loop part of rms displacement comes from scales typically considered to be
nonlinear ($k>0.2{h/ \rm Mpc}$), while the linear part is mostly determined by $k<0.2{h/ \rm Mpc}$. 
As a result, we can trust the linear prediction better than the one loop part. There is a small part of one loop contribution 
that comes from $k<0.2{h/ \rm Mpc}$,
which we can reliably compute, and which adds to the linear rms displacement while most of the one loop part is suppressed by the 
halo formation. There is also a small part of linear displacement that comes from $k>0.2{h/ \rm Mpc}$, and is suppressed by 
the halo formation just as its one loop counterpart. Relative to the linear displacement, we thus have one positive correction 
from one loop LPT (mostly at $k<0.2{h/ \rm Mpc}$), and one correction that reduces the linear displacement for $k>0.2{h/ \rm Mpc}$, 
and that is effectively negative relative to linear value. It appears the two cancel each other so that the total is very close to a 
linear value to a level of 1-1.5\%, almost independent of redshift.

We note in passing that this has an implication to the effective field theory (EFT) of large scale structure approach 
\cite{Carrasco:2012cv, Porto:2013qua}. In Eulerian approach the high $k$ part, which PT cannot reliably evaluate, 
is parametrized with a free parameter $\alpha$ in the $\alpha k^2 P_{\rm lin}$ term. This term is then added to 
the one loop PT power spectrum result. The value of this correction obtained from fits to the simulation power 
spectrum is estimated to be of order of 10\% of $\sigma^2$ at $z=0$ \cite{Carrasco:2012cv}. As shown earlier 
(equation \eqref{eq:spt}), Eulerian SPT can be obtained from expanding LPT at given order (in the SPT sense). At 
one loop level, leading low k corrections $\sim k^2P_{\rm lin}$ comes from combining several terms in LPT: $R_1$, 
$R_2$ and $\sigma^2 k^2P_{\rm lin}$. For the displacement field dispersion $\sigma^2$, we have seen above that 
the nonlinear correction to linear value has to be very small. In order to compute the 
low k Eulerian EFT results, leading corrections to the terms like $R_1$ and $R_2$ from LPT also need to be computed. 
This offers, thus, an independent consistency check of EFT approach in Lagrangian and Eulerian framework. To perform 
this check, EFT corrections to the two, three and possibly four point displacement cumulants need to be calibrated from 
the simulation measurements of these cumulants. 

Since the decomposition of the power spectrum in equation \eqref{eq:PSmodel} is additive in all of its  
contributions, it follows for the total correlation function:
\begin{align}
\xi(r) = \xi_{\text{Zel}}(r)+\xi_{W}(r)+\xi_{BB} (r),
\end{align}
where each of the constituents is respectively the Fourier transform of the terms 
in \eqref{eq:PSmodel}.

\begin{figure*}[!t]
    \centering
    \includegraphics[trim = 3.0cm 0cm 0cm 0cm, clip, scale=0.45]{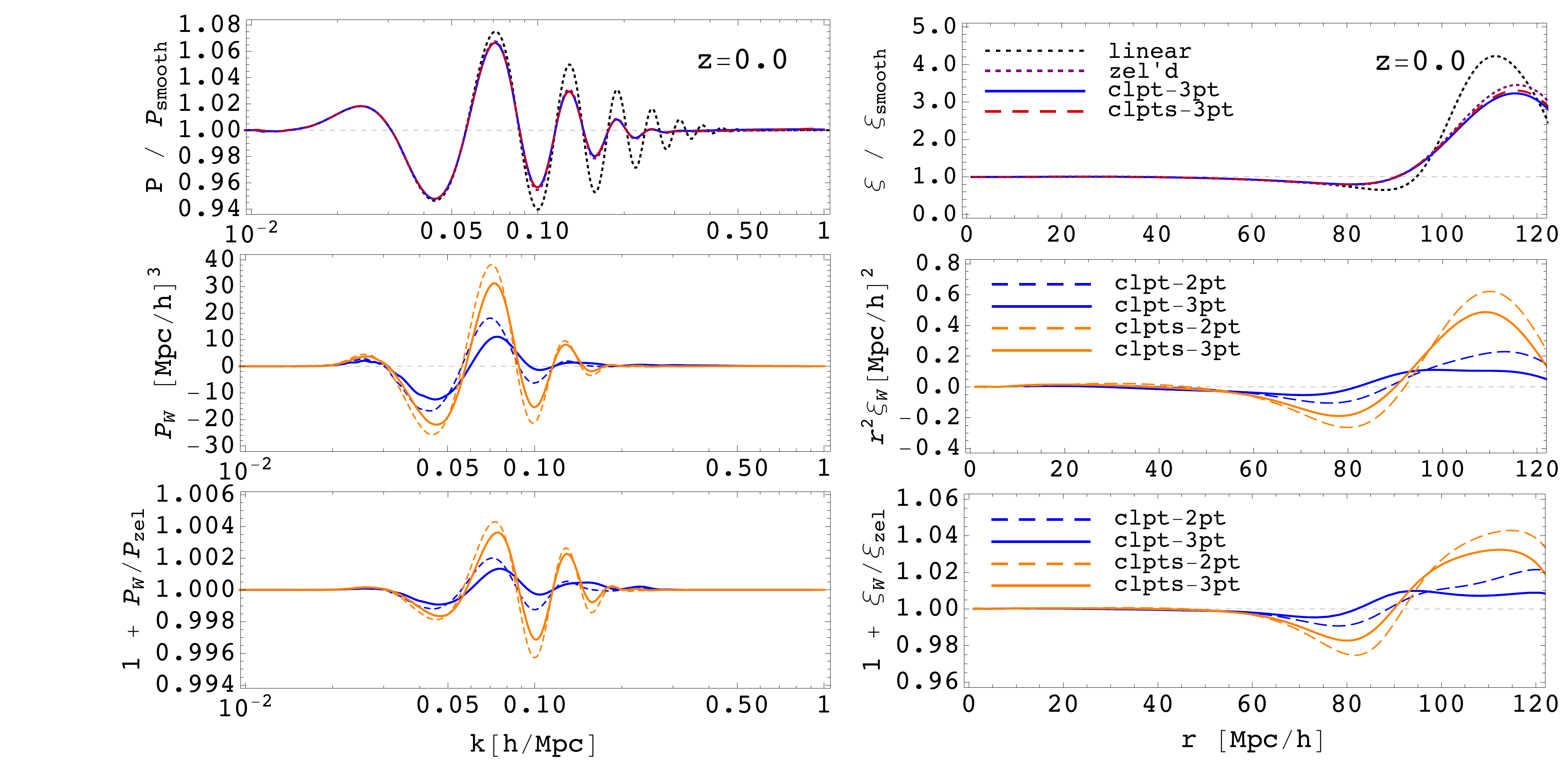}
    \caption{\small \textit{Upper panels:} On the left panel, we show the power spectrum 
    divided by the corresponding no-wiggle power spectrum, which was constructed to give the limits of 1 at both low and 
    high k for all the lines. We show the results for the linear theory (black dotted line), 
    Zel'dovich (purple dotted line), one loop CLPT (blue solid line) and CLPTs (red dashed line).
    On the right panel, we show the same results for the correlation function divided by the 
    corresponding no-wiggle version.\\ 
    \textit{Middle panels:} On the left panel, we show the residual wiggle power spectrum $P_W$:
    one loop CLPT-2pt (blue dashed line) and CLPTs-2pt (orange dashed line) results do not include the 
    bispectum terms $V$ and $T$, and one loop CLPT-3pt (blue solid line) and CLPTs-3pt (orange solid line) 
    results do include $V$ and $T$ terms. On the right panel we show the corresponding results in the 
    correlation function $r^2\xi_W$.\\
     \textit{Bottom panels:} The same lines are shown as in the middle panels divided by the Zel'dovich
    result in order to highlight the relative effects.
    }
    \label{fig:PWplot}
\end{figure*}

We investigate the consequence of adopting a simple Lorentzian like damping 
of the displacement field power spectrum suggested in \cite{Chan:2013vao}, 
$p(k) \rightarrow p(k)/(1+\alpha k^n)$, 
where $\alpha$ and $n$ are free parameters evaluated in simulations (see \cite{Chan:2013vao}
for numerical values). 
The correction above 
also changes the leading low $k$ dependence of the displacement spectrum. Since there is no
reason to do this we apply this correction only at the scales higher 
than $k=0.1h/$Mpc, and smoothly interpolate to no correction below that $k$. 
Using this result, we correct the $X$ and $Y$ functions, but not $V$ and $T$ which would require separate N-body analysis of a 
displacement bispectrum. 
We call this result CLPTs.
As shown on the right panel of figure \ref{fig:Sigmaplot} this reduces the rms displacement variance $\sigma^2$ to a value close 
to linear. 

\subsection{BAO wiggles: beyond CLPT}

Let us focus first on the residual wiggle part, $P_{W}$, for which we can use the CLPT and CLPTs results discussed above. 
The key in extracting just the wiggle information is in the construction of a reliable no-wiggle power spectrum. 
In order to achieve this we use the b-spline smoothing method similar to the one used in \cite{Reid:2009xm}. 
We first obtain the smooth version of the linear power spectrum by smoothing a realistic linear power spectrum.  
This is most easily achieved when the linear power spectrum is first divided by the no-wiggle fitting spectra from 
\cite{Eisenstein:1997ik}, then smoothed, and then multiplied again with the no-wiggle fitting spectra. 
We then use this smoothed linear power spectrum as input to compute Zel'dovich and one loop CLPT result. 
Since there is no unique way in obtaining the smoothed line we construct a family of smooth approximations 
and construct the final result as a linear composition of those. Further on we impose two integral constraint requirements, 
i.e. we want $\sigma_8$ and $\sigma_v$ to be the same in the case of smoothed spectrum and the one with BAO wiggles. 
This guarantees that any no-wiggle nonlinear power spectrum obtained in this way will agree well with the spectrum 
containing wiggles even for high k values. Using this smoothing on Zel'dovich and one loop CLPT result we 
extract the residual wiggle spectrum
\begin{align}
P_W(k) = (P_{CLPT} - P_{Zel} ) - (P_{CLPT} - P_{Zel} )_{nw},
\label{eq:decPS}
\end{align}
where $P_{CLPT}$ is the one loop CLPT power spectrum and $P_{Zel}$ is the Zel'dovich one. Subscript 
$``nw"$ stands for the no-wiggle spectrum, i.e. the one with smoothed BAO wiggles. 
In order to obtain the wiggles results for the CLPTs we just replace the CLPT power spectrum with the 
corresponding CLPTs power spectrum in equation \eqref{eq:decPS} above.

The wiggle results obtained by this procedure are shown in figure \ref{fig:PWplot}.  
In the upper panel we show the results for the power spectrum divided by the corresponding smooth
spectrum (so e.g. linear divided by the smooth linear, Zel'dovich by the smooth Zel'dovich etc.), and similar for 
the correlation function. We notice that while CLPT and CLPTs are almost indistinguishable in the power spectrum 
in the correlation function one can notice slight differences.  In order to see these effects better, in the middle 
panels of the same figure we show the individual $P_W$ and $\xi_W$ contributions, and in the bottom panels the 
same wiggle contribution relative to the Zel'dovich result. 
We see that the BAO correlation function feature does not change much around BAO,  which is in agreement with
the one loop CLPT result presented in figure \ref{fig:CorrFunction}. One can see that the nonlinear
effects beyond Zel'dovich are at a level of 1\%  in CLPT case and 2-3\% in CLPTs in the correlation function, while in 
the power spectrum they appear to be 10 times smaller (~0.1-0.3\%)  in the amplitude compared to the Zel'dovich 
power spectrum. As seen in figure \ref{fig:CorrFunction}, and later on in figure \ref{fig:last}, simulation measurements 
of the correlation function tend to favor the CLPTs  wiggle results. We emphasize however that we are exploring 
very small effects, and it is not clear whether these wiggle effects beyond Zel'dovich can even be observable: the deviations
are of order of a few percent in amplitude in the correlation function at $100$Mpc/$h$, where the sampling variance errors are very 
large. In the power spectrum the effects appear even smaller in amplitude although less localized. 
We also show the results where $V$ and $T$ terms are neglected which has the effect of slightly raising the amplitude of 
wiggles in both CLPT and CLPTs case. We can thus say that $V$ and $T$ contribute to the smoothing of the BAO wiggles. 
We conclude that CLPTs is probably an improvement over one loop CLPT for wiggles, but the effects are small when compared to the 
dominant Zel'dovich effects.  

\subsection{Broadband power: beyond CLPT}

\begin{figure*}[!t]
    \centering
    \includegraphics[scale=0.4]{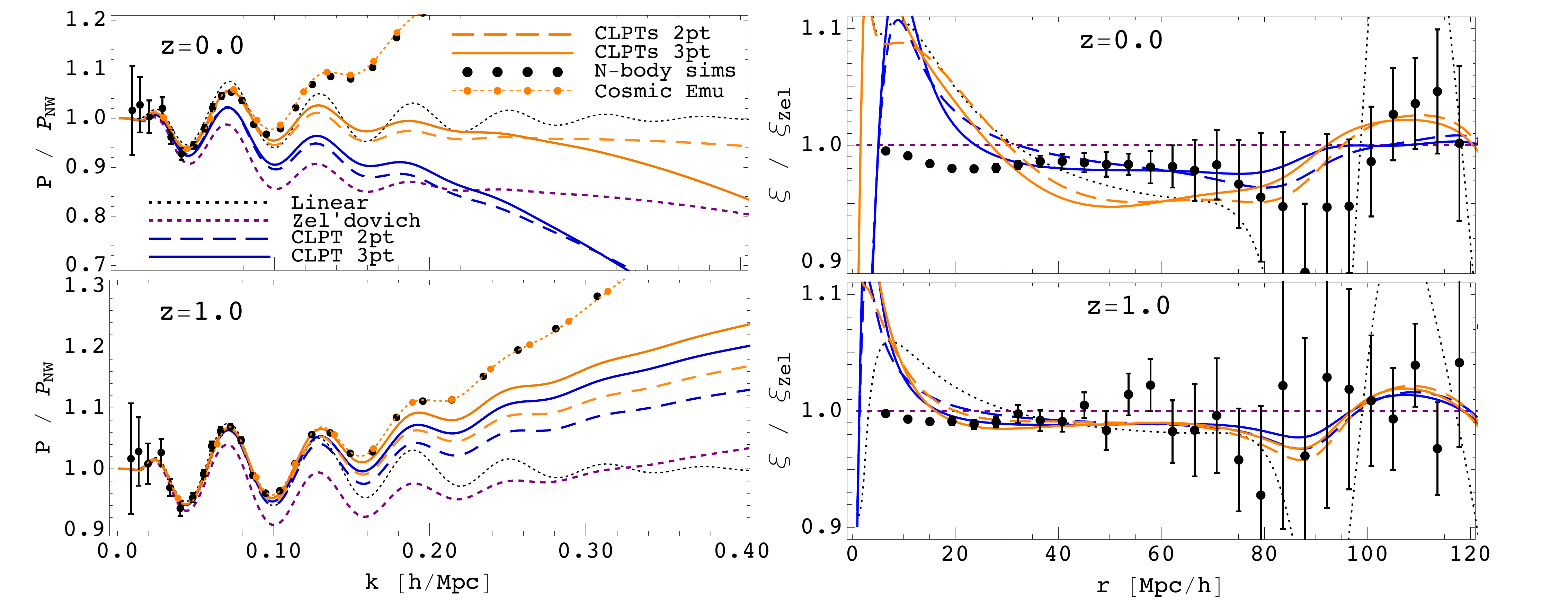}
    \caption{\small (\textit{Left panels:}) 
    Comparison of the power spectrum using CLPT (blue lines) and CLPTs (orange lines) at redshift $z=0.0$ and $z=1.0$.    
    We show the contributions in both cases when three point terms $V$ and $T$ are neglected (dashed blue and orange lines) 
    and taken into account using CLPT prediction for the 3-point function 
(solid blue and orange lines). We show also the results of linear theory (black dotted line), Zel'dovich
    approximation (purple solid line) N-body simulation (black dots) and Cosmic emulator (orange dots). All the lines are divided 
    by the no-wiggle linear theory. 
    (\textit{Right panels:}) Comparison of the correlation function using CLPT (blue lines) and CLPTs (orange lines) at redshift $z=0.0$ 
    and $z=1.0$. Color labeling is the same as for the left panels. Results are divided by the Zel'dovich correlation function.
    }
    \label{fig:last}
\end{figure*}

As discussed in the previous section, one loop CLPT is a poor model for the power spectrum although it improves upon 
Zel'dovich. In figure \ref{fig:last} we show the power spectrum and correlation function results comparing CLPTs and CLPT. 
We find that CLPTs improves upon CLPT in the power spectrum, but still fails to reach a good agreement with the  
N-body simulations. This suggests that modeling of the $X$ and $Y$ term by CLPTs introduces positive change but 
is not sufficient and similar procedure would need to be preformed on $V$ and $T$ terms and possibly also higher cumulants.
Physically, both Zel'dovich and one loop CLPT fail to make halos, they can only create the
onset of halo formation. CLPT is unable to adhese dark matter particles inside the halos. This is a well known 
problem for Zel'dovich approximation, where the particles simply stream straight on their trajectories set by initial 
conditions. Because the halo formation is missing one cannot expect CLPT to do well in the power spectrum, where 
one halo terms are a dominant contribution to the power spectrum already at a relatively low $k$ \cite{Mohammed:2014lja}.

In the same figure we also show the correlation function results for one loop CLPT, which achieves a good agreement with simulations 
above $30$Mpc/h. Interestingly, this is not improved by CLPTs, which performs worse on the broad band part $\xi_{BB}$
than CLPT. Below $30$Mpc/h however both one loop CLPT  and CLPTs are worse than Zel'dovich. CLPT and CLPTs are therefore not 
an obvious improvement over Zel'dovich for the correlation function. 

\section{Conclusions}
\label{sec:conc}

In this paper we use the Lagrangian perturbation theory for evaluation of the displacement fields in order to compute up to one 
loop contributions to the cumulants in the resulting power spectrum. 
We start by reviewing the Lagrangian framework of describing the overdensity of dark matter particles and present the framework 
of computing the two point function in Fourier space, i.e. the power spectrum. In this framework (first suggested in \cite{Carlson:2012bu} 
for the correlation function) power spectrum is given as a sum of n-point cumulants of the difference of the displacement field 
in two points $\Delta=\Psi_2-\Psi_1$. In this form, the translation invariance is given explicitly in each of the cumulants 
(see also \cite{Sugiyama:2013mpa}).
Only the first two cumulants (two-point and three-point) have contributions at one loop and we show that scale dependence 
of each of these. Computationally difficult part of integrating a highly oscillatory integrand is overcome by angular 
moment expansion. This yields the result expressed as the sum of integrals with spherical Bassel functions, which 
ensure the quick convergence rate of the sum. Alternative scheme of solving these equations by expanding the 
integrand into spherical harmonics gives identical results. We note that the convergence rate of the first expansion 
method is somewhat better since it involves a partial resummation of some of the terms. We compare the final result to 
the N-body simulations. 
A similar analysis to ours has been presented in \cite{Sugiyama:2013mpa} and
wherever it is comparable, the results agree.
We Fourier transform our result for the one loop power spectrum to obtain the correlation function predictions.
The CLPT results found in this paper, when 
Fourier transformed, agree with \cite{Carlson:2012bu}. 

For low redshifts, our power spectrum results do not agree well with the N-body simulations.  We argue that 
one of the main reasons is that perturbation theory for the displacement field overpredicts the rms displacement at small scales. 
A possible improvement is to implement the suppression of the
displacement field power spectrum, motivated by simulations (\cite{Chan:2013vao}), which we call CLPTs. 
This offers a consistent treatment of the two point displacement 
functions $X$ and $Y$. We evaluate the rms displacement $\sigma_v$ in simulations 
and compare it to this model, finding a good agreement. 
We note that the numerical value of the zero lag rms displacement is nearly identical to the linear value: 
this appears to be a result of a cancellation between a small positive 1 loop contribution and the high $k$ 
suppression of the small linear contribution. 
In the absence of any available guidance from simulations we do not modify
the tree point functions $V$ and $T$. The result of this procedure improves the overall 
broad band behaviour of the power spectrum, but still fails to reproduce the final N-body 
simulations result. Neither CLPT nor CLPTs has a large effect on BAO wiggles in the power spectrum. 

We look at the cross and auto spectrum at 2LPT level, i.e. up to the second order expansion of the displacement field. In this 
way, we can directly compare the performance of our analytic solution with the measurements of the spectrum on the grid 
obtained by displacing the particles using initial conditions code \cite{Crocce:2006ve}. We compare six different power 
spectra: three auto spectra, Zel'dovich, 2LPT (second order only) and Zel'dovich+2LPT, and similarly three cross spectra. 
We find a very good agreement of the analytic and measured predictions in all of the spectra. Differences that are noticeable 
as we approach higher scales ($k\sim0.4{h/ \rm Mpc}$) are due to resolution effects of grid measurements which start to affect the results. 
In addition, even though both calculations are of the same perturbative order in the displacement field, in our analytic approach 
we truncate the cumulant expansion at one loop order, while grid measurements in principle have contributions from higher 
orders of cumulant expansion for some of the spectra. These differences observed between the analytic solutions and solution 
on the grid are significantly smaller than what we have seen earlier in the fully nonlinear case when comparing to the full N-body 
simulations. This suggests that truncating the cumulant expansion leads to a very good approximation of the nonlinear power spectrum.
More accurate modeling of the analytic power spectrum thus requires a more accurate modeling of the displacement field spectra 
in the cumulant expansion (i.e. X, Y, V and T terms).  

While for the power spectrum all of the models lack power compared to simulations, with power increasing from Zel'dovich to CLPT to CLPTs,
in the correlation function the picture is that 
Zel'dovich approximation is a remarkably accurate model of the correlation function (see also \cite{White:2014gfa}), with 
deviations from simulations at a few 
percent level for $r>5$Mpc/h, especially at higher redshifts. CLPT and CLPTs do not significantly improve upon it, and 
in fact are worse that Zel'dovich at smaller radii. 
There are residual BAO wiggle effects beyond Zel'dovich at a level of a few percent: these appear to be improved 
by CLPT, and improved further by CLPTs. These effects are at a few percent level on 100Mpc/h scale, so 
it is unclear how observable they are, given the large sampling variance fluctuations in a realistic survey. 
We conclude that Zel'dovich, CLPT and its CLPTs extension give very different results in the power spectrum versus correlation 
function, and neither of them agrees with simulations in all aspects. CLPTs uses (approximately) exact two point correlator 
of the displacement field, so the fact that it still gives inaccurate results implies the higher order displacement field 
correlators play a role, starting from the V and T terms. 
A model that would give correct results both in the power spectrum and in the correlation function therefore remains elusive. 

\section*{Acknowledgments}

We would like to thank Raul Angulo, Teppei Okumura, Leonardo Senatore, Marcel Schmittfull, Martin White and Jaiyul Yoo 
for useful discussions and comments. ZV would like to thank the Berkeley Center for Cosmological Physics and the Lawrence 
Berkeley Laboratory for their hospitality. ZV would also like to thank Thomas Gehrmann and Nenad Smontara for stimulating 
discussion related to technical aspects of this work. This work is supported by 
NASA ATP grant NNX12AG71G and University Zurich under the grant FK-13-102. 
T.B. gratefully acknowledges support from the Institute for Advanced Study through the W.~M.~Keck Foundation Fund.
We use CUBA libery \cite{Hahn:2004fe} for numerical integration, and FFTLog \cite{Hamilton:1999uv} for the integrals 
involving sperical Bessel funcitons. The simulations were performed on the ZBOX4 supercomputer of the Institute 
for Computational Science at the University of Z\"{u}rich.


\appendix

\section{PT computation for $A_{ij}$ term}
\label{sec:XY}

In this section we show the one loop LPT calculation of the $A_{ij} = \langle \Delta_i \Delta_j \rangle_c$, a similar 
result can be found in \cite{Carlson:2012bu}. From the definition of $\Delta$ we have
\begin{align}
  \Delta_i &= \Psi_i(\VEC{q}_2) - \Psi_i(\VEC{q}_1)  \nonumber\\
 &=\int \frac{d^3p}{(2\pi)^3}\left(e^{-i\VEC{p}\cdot\VEC{q}_2} - e^{-i\VEC{p}\cdot\VEC{q}_1}\right) \Psi_i(\VEC{p}).
\end{align}
From equation \eqref{eq:npsi} the two point function is given by
\begin{equation}
    \label{eq:CM}
    \left\langle \tilde{\Psi}_i(\VEC{p}_1) \tilde{\Psi}_j(\VEC{p}_2) \right\rangle_c
        = (2\pi)^3 \delta^D(\VEC{p}_1 + \VEC{p}_2) C_{ij}(\VEC{p}_1),
\end{equation}
where the $C_{ij}(\VEC{k})$ are the two point displacement power spectra. One loop LPT
prediction for these spectra gives the contributions 
\begin{gather}
  C_{ij}^{(11)}(\VEC{k}) = \frac{k_i k_j}{k^4} P_L(k) \\
  C_{ij}^{(22)}(\VEC{k}) = \frac{9}{98} \frac{k_i k_j}{k^4} Q_1(k)  \\
  C_{ij}^{(13)}(\VEC{k}) = C_{ij}^{(31)}(\VEC{k}) = \frac{5}{21}
        \frac{k_i k_j}{k^4} R_1(k).
\end{gather}
where $R_n$ and $Q_n$ terms are defined as in \cite{Matsubara:2007wj,Carlson:2012bu}.
We can simplify this result by writing $C_{ij} = (k_i k_j/k^4)p(k)$, where $p(k)$ 
is the one loop displacement spectra defined in equation \eqref{eq:malip}. We have
\begin{align}
  A_{ij} =& 2\int \frac{d^3k}{(2\pi)^3}
  \big(1 - \cos\left(\VEC{k}\cdot\VEC{q}\big)\right)\nonumber\\
  &\times \frac{k_i k_j}{k^4} \left(P_L(k)+\frac{9}{98} Q_1(k)+\frac{10}{21} R_1(k)\right) .
\end{align}
Contracting this quantity first by $\delta_{ij}$ and then by $\hat{q}_i \hat{q}_j$, we
obtain the system of equations
\begin{align}
\left.
\begin{array}{ll}
A_0\equiv\df^K_{ij}A_{ij}=3X+Y\\
\bar{A}\equiv\hat{q}_i\hat{q}_jA_{ij}=X+Y
\end{array} \right\rbrace
\rightarrow
\begin{array}{rl}
X&=\frac{1}{2}(A_0-\bar{A})\\
Y&=\frac{1}{2}(3\bar{A}-A_0),
\end{array}
\label{eq:XYdef} 
\end{align}
Defining the $\mu = \hat{k}\cdot\hat{q}$ and by perform the angular integrations,
we get
\begin{align}
  X(q) =& \int_0^\infty \frac{dk}{\pi^2} \left(P_L(k)+\frac{9}{98} Q_1(k)+\frac{10}{21} R_1(k)\right)\nonumber\\
  &\times \left(\frac{1}{3} - \frac{j_1(kq)}{kq}\right) , \\
  Y(q) =& \int_0^\infty \frac{dk}{\pi^2} \left(P_L(k)+\frac{9}{98} Q_1(k)+\frac{10}{21} R_1(k)\right)\nonumber\\
  &\times j_2(kq).
  \label{eq:XYex}
\end{align}
In small $q$ limit $X$ and $Y$ vanish while in a large $q$ limit $Y\rightarrow0$
and $X\rightarrow \sigma^2$, where $\sigma$ is defined in equation \eqref{eq:sig}.
Full $q$ dependence of $X$ and $Y$ terms at $z=0$ is shown in the left panel of figure \ref{fig:XYTV}.

\section{PT computation for $W_{ijk}$ term}
\label{sec:VT}

As in the previous section using the Fourier transform of the field $\Psi(\VEC{q})$
we have 
\begin{align}
W_{ijk}(\VEC{q})=&\int \prod_{n=1}^3 \frac{d^3p_n}{(2\pi)^3}
\left(e^{-i\VEC{q}_2\cdot\VEC{p}_n}-e^{-i\VEC{q}_1\cdot\VEC{p}_n}\right)\nonumber\\
&\times\left\langle \Psi_i(\VEC{p}_1)\Psi_j(\VEC{p}_2)\Psi_k(\VEC{p}_3)\right\rangle\nonumber\\
=&i(2\pi)^3\int \prod_{n=1}^3 \frac{d^3p_n}{(2\pi)^3}
\left(e^{-i\VEC{q}_2\cdot\VEC{p}_n}-e^{-i\VEC{q}_1\cdot\VEC{p}_n}\right)\nonumber\\
&\times\df^D_{123}C_{ijk}(\VEC{p}_1,\VEC{p}_2,\VEC{p}_3)\nonumber\\
=&i(2\pi)^3\df\dbinom{ijk}{nmr}\nonumber\\
&\times\int\prod_{n=1}^3\frac{d^3p_n}{(2\pi)^3}
\left(e^{-i\VEC{q}_2\cdot\VEC{p}_n}-e^{-i\VEC{q}_1\cdot\VEC{p}_n}\right)\nonumber\\
&\times\df^D_{123}C^{(112)}_{nmr}(\VEC{p}_1,\VEC{p}_2,\VEC{p}_3),
\end{align}
where in the last line we have expanded $C_{ijk}$ in terms of the one loop contributions and symmetrised the indexes. 
We use the abbreviation for the Dirac delta function $\df^D_{123}=\df^D(\VEC{p}_1+\VEC{p}_2+\VEC{p}_3)$, as well as 
for the symmetrized sum of Kronecker deltas
\begin{align}
\df\dbinom{ijk}{nmr}=\df^K_{in}\df^K_{jm}\df^K_{kr}+
\df^K_{kn}\df^K_{im}\df^K_{jr}+\df^K_{jn}\df^K_{km}\df^K_{ir}.
\end{align}
Contracting the tensor we get 
\begin{align}
\overline{W}(q)=&3i(2\pi)^3  \int \prod_{n=1}^3 \frac{d^3p_n}{(2\pi)^3}
\left(e^{-i\VEC{q}_2\cdot\VEC{p}_n}-e^{-i\VEC{q}_1\cdot\VEC{p}_n}\right)\df^D_{123}\nonumber\\
&\times\hat{q}_i\hat{q}_j\hat{q}_kC_{ijk}^{(112)}(\VEC{p}_1,\VEC{p}_2,\VEC{p}_3),\nonumber\\
W_0(q)=&i(2\pi)^3 \int \prod_{n=1}^3 \frac{d^3p_n}{(2\pi)^3}
\left(e^{-i\VEC{q}_2\cdot\VEC{p}_n}-e^{i\VEC{q}_1\cdot\VEC{p}_n}\right)\df^D_{123}\nonumber\\
&\times\Big(2\hat{q}_iC_{ijj}^{(112)}(\VEC{p}_1,\VEC{p}_2,\VEC{p}_3)
+\hat{q}_iC_{jji}^{(112)}(\VEC{p}_1,\VEC{p}_2,\VEC{p}_3)\Big).
\end{align}
Before we evaluate these integrals one by one let us first do some general simplifications. 
First we use the delta function to integrate out the $\VEC{p}_3$ momentum which gives 
\begin{align}
i(2\pi)^3 \int &\prod_{n=1}^3 \frac{d^3p_n}{(2\pi)^3}\left(e^{-i\VEC{q}_2\cdot\VEC{p}_n}
-e^{-i\VEC{q}_1\cdot\VEC{p}_n}\right)\df^D_{123}\nonumber\\
&\times C_{ijk}(\VEC{p}_1,\VEC{p}_2,\VEC{p}_3)=\nonumber\\
2\int \frac{d^3p_1}{(2\pi)^3}&\frac{d^3p_2}{(2\pi)^3}\Big(\sin\left(\VEC{q}\cdot\VEC{p}_1\right)
+\sin (\VEC{q}\cdot\VEC{p}_2)\nonumber\\
&-\sin\big(\VEC{q}\cdot(\VEC{p}_1+\VEC{p}_2)\big)\Big)C_{ijk}(\VEC{p}_1,\VEC{p}_2,\VEC{p}_3).
\end{align}
After some straightforward computation these integrals can be written in the form
\begin{align}
\overline{W}(q)=&\frac{6}{5}\int \frac{dk}{2{\pi}^2 k}\left(-\frac{3}{7}\right)\bigg(Q_1(k)-3Q_2(k)\nonumber\\
&\qquad \qquad +2R_1(k)-6R_2(k)\bigg)j_1(qk)\nonumber\\
&+\frac{6}{5}\int \frac{dk}{2{\pi}^2 k}\left(-\frac{3}{7}\right)\bigg(Q_1(k)+2Q_2(k)\nonumber\\
&\qquad\qquad  +2R_1(k)+4R_2(k)\bigg)j_3(qk),
\end{align}
and
\begin{align}
W_0(q)=&2\int \frac{dk}{2{\pi}^2 k}\left(-\frac{3}{7}\right)\big(Q_1(k)-3Q_2(k)\nonumber\\
&\qquad\qquad +2R_1(k)-6R_2(k)\big)j_1(qk),
\end{align}
where $R_n$ and $Q_n$ terms are defined as in \cite{Matsubara:2007wj, Carlson:2012bu}. Using the 
transformations from equation \eqref{eq:TVdef} we get the final one loop estimate of the displacement 
bispectrum contribution to the density power spectrum
\begin{align}
T(q)&=3\int \frac{dk}{2{\pi}^2 k}\left(-\frac{3}{7}\right)\Big(Q_1(k)+2Q_2(k)\nonumber\\
&\qquad\qquad +2R_1(k)+4R_2(k)\Big)j_3(qk),\nonumber\\
V(q)&=\frac{1}{5}\big(W_0(q)-T(q)\big).
\label{eq:VTex}
\end{align}
Note that each of these quantities approaches 0 as $q \to 0$ as well as in $q\to \infty$ limit.
Full $q$ dependence of $V$ and $T$ terms at $z=0$ is shown in the right panel in figure \ref{fig:XYTV}.

\section{Angular integration: Method I}
\label{sec:int1}

As shown in \cite{Schneider:1995} the angular integral that appears in Zel'dovich limit can be expressed as
\begin{align}
\int^1_{-1} d\mu~e^{iA\mu}~\exp(B\mu^2)=2e^B \sum_{n=0}^\infty \left(-\frac{2B}{A}\right)^nj_n(A),
\label{eq:intAB}
\end{align}
where $j_n$ are the spherical Bessel functions. We can define the $k-$th moment of the integral function $M_k$ as
\begin{align}
M_k(A,B) = \int_{-1}^1 d \mu ~ \mu^k ~e^{iA\mu}~\exp(B\mu^2).
\end{align}
If we take the $k$-derivative of this integral with respect to $A$ we get the expression for the $k-$th moment 
\begin{align}
M_k(A,B)=2~(-i)^k e^B \sum_{n=0}^\infty \left(-2B\right)^n\left(\frac{d}{dA}\right)^k A^{-n} j_n(A),
\end{align}
where we can use the relation for the spherical Bessel functions
\begin{align}
\left(\frac{1}{\nu}\frac{d}{d\nu}\right)^k\left(\nu^{-n} j_n(\nu)\right)=(-1)^k \nu^{-n-k}j_{n+k}.
\end{align}
Finally, we are interested in the case where we have $i\epsilon\mu^3$ term in the exponent.
Expanding the left hand side in $\epsilon$ we get
\begin{align}
\int^1_{-1} d\mu~ e^{iA\mu}~\exp(B&\mu^2 + i\epsilon\mu^3)
=\sum_{l=0}^\infty \frac{(i\epsilon)^l}{l!} M_{3l}(A,B)\nonumber\\
&=2 e^B \sum_{n=0}^\infty \left(-\frac{2B}{A}\right)^n J_n(A,\epsilon),
\label{eq:epsint}
\end{align}
where we have defined a new function $J_n$ which deviates from the spherical Bessel function $j_n$
depending on the values of parameter $\epsilon$, and is given in the form of a series
\begin{align}
 J_n(A,\epsilon)=\sum_{l=0}^\infty &(-1)^l\frac{\epsilon^{2l}}{(2A)^{3l}}\Big[F_1^{(n,l)}(A)\nonumber\\
&-3(3l+1)(6l+1)\frac{\epsilon}{A}F_2^{(n,l)}(A)\Big],
\label{eq:newJ}
\end{align}
with the additional auxiliary functions defined as
\begin{align}
F_1^{(n,l)}(A)&=\frac{(6l)!}{(2l)!}\sum_{p=0}^{3l}\frac{(-2A)^p}{(2p)!(3l-p)!}j_{n+3l+p}(A),\nonumber\\
F_2^{(n,l)}(A)&=\frac{(6l)!}{(2l)!}\sum_{p=0}^{3l+1}\frac{(-2A)^p}{(2p+1)!(3l-p+1)!}j_{n+3l+p+2}(A).
\label{eq:ffnctions}
\end{align}
We can approximate the result at a first order in $\epsilon$, which would correspond to the
solution presented in \cite{Carlson:2012bu} for the correlation function. We have 
\begin{align}
\int^1_{-1} &d\mu~ e^{iA\mu+B\mu^2}(1+i\epsilon\mu^3)=\nonumber\\
&=2 e^B \sum_{n=0}^\infty \left(\frac{-2B}{A}\right)^n 
\left[F_1^{(n,0)}(A)-3\frac{\epsilon}{A}F_2^{(n,0)}(A)\right]\nonumber\\
&=2 e^B \sum_{n=0}^\infty \left(\frac{-2B}{A}\right)^n \Big(j_n(A)-\frac{\epsilon}{A}\big[3j_{n+2}(A)\nonumber\\
&\qquad\qquad\qquad\qquad\qquad-Aj_{n+3}(A)\big]\Big).
\end{align}
In the limit $\epsilon\rightarrow 0$ it is clear that we regain the old result in equation \eqref{eq:intAB}.
If we compare our full integral \eqref{eq:PSwithVT} to the equation \eqref{eq:epsint} we get the correspondence
\begin{align}
A(k,q)&=k\left(q-\frac{1}{2}k^2V(q)\right),\nonumber\\
B(k,q)&=-\frac{1}{2}k^2Y(q),\nonumber\\
\epsilon(k,q)&=-\frac{1}{6}k^3T(q).
\label{eq:ABee}
\end{align}
From equation \eqref{eq:epsint} and \eqref{eq:intAB} we see that the expansion parameters of the series 
are $-2B/A$, $-\epsilon^2$, and $-\epsilon/A$. We find that the bispectrum terms $V$ and $T$ start to be 
relevant for higher $k$ as one would expect, as we see from low-$k$ calculation in section \ref{sec:lowk}.

\section{Angular integration: Method II}
\label{sec:int2}

\begin{figure*}[!t]
    \centering
    \includegraphics[trim = 0.95cm 0cm 0cm 0cm, clip,scale=0.558]{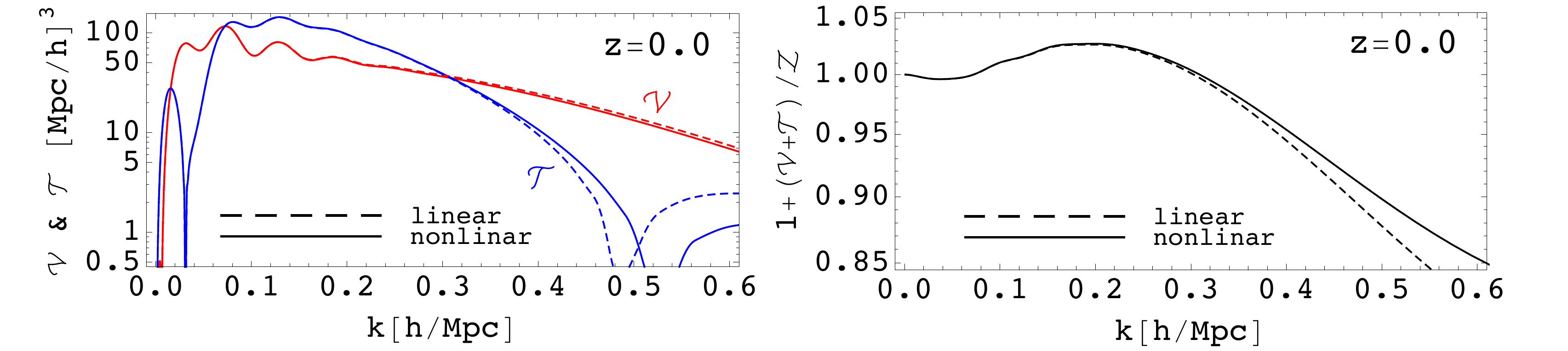}
    \caption{\small (\textit{Left panel:})
    Terms $\mathcal{V}$ (red lines) and $\mathcal{T}$ (blue lines), from equation \eqref{eq:ZVT}. Full results (solid lines) are obtained 
    using method I or II described in sections \ref{sec:int1} and \ref{sec:int2}. In linearized version (dashed line) only the leading contribution 
    of $V$ and $T$ terms is kept. 
    (\textit{Right panel:}) Relative contribution of the sum $\mathcal{V}+\mathcal{T}$ to the two point contribution $\mathcal{Z}$ is shown in 
    the fully nonlinear case (solid lines) and in the linearized version (dashed lines).
    Results are shown at redshift $z=0.0$.    
    }
    \label{fig:VT}
\end{figure*}

In this section we present an alternative derivation of the solution of the equation \eqref{eq:epsint} integral. The idea is to generalize 
the plane wave expansion (for review see e.g. \cite{Mehrem:2009}) in order to obtain the solution of the integral.
We start from a well known formula of the plane wave expansion
\begin{align}
e^{i \VEC{k}\cdot\VEC{r}}=\sum_{l=0}^\infty i^l(2l+1)P_l(\cos\theta)j_l(kr),
\label{eq:planewave}
\end{align}
where  $\theta$ is the angle between $\VEC{k}$ and $\VEC{r}$. Expanding this, it follows that for the powers of 
the plane wave variable we have
\begin{align}
(i x)^n &=\left( \frac{d^n}{d\alpha^n}e^{i\alpha x}\right)_{\alpha=0}\nonumber\\
&=\sum_{l=0}^\infty i^l(2l+1)P_l(x)\left(\frac{d^nj_l(\alpha)}{d\alpha^n}\right)_{\alpha=0}.
\end{align}
This gives us the Taylor expansion of the spherical Bessel function around the zero
\begin{align}
j_l(\alpha)=\sum_{n=0}^\infty\frac{\alpha^n}{n!}\left(\frac{d^nj_l(\alpha)}{d\alpha^n}\right)_{\alpha=0}.
\end{align}
We can compare this expansion to the well known form of the series representation of spherical Bessel functions 
(for reference see e.g. \cite{NIST:DLMF,Olver:2010:NHMF})
\begin{align}
j_l(\alpha)=\alpha^l\sum_{k=0}^\infty\frac{(-)^k}{2^k k!}\frac{\alpha^{2k}}{(2l+2k+1)!!},
\end{align}
which gives the coefficients of the Taylor expression above
\begin{align}
b_n^l&=\left(\frac{d^nj_l(\alpha)}{d\alpha^n}\right)_{\alpha=0}\nonumber\\
&=\begin{cases} 
\frac{i^{n-l} n!}{\sqrt{2}^{n-l}\left(\frac{1}{2}(n-l)\right)!(n+l+1)!!}, & \small {\mbox{if } n\geq l ~\&~ n\mbox{ and }}
\\ & \small{l\mbox{ both even or odd,} }
\\\qquad\qquad 0, & \mbox{otherwise.}
\end{cases}
\end{align}
Using these coefficients we have
\begin{align}
(i x)^n =\sum_{l=0}^\infty i^l(2l+1)P_l(x)b_n^l,
\end{align}
from which follows 
\begin{align}
e^{Bx^2}&=\sum_{l=0}^\infty i^l(2l+1)P_l(x)\sum_{n=0}^\infty \frac{(-B)^n}{n!}b_{2n}^l,\nonumber\\
e^{i\epsilon x^3}&=\sum_{l=0}^\infty i^l(2l+1)P_l(x)\sum_{n=0}^\infty \frac{(-\epsilon)^n}{n!}b_{3n}^l.
\end{align}
After some calculation we have the solution of the integral \eqref{eq:intAB} in form of the series
\begin{align}
\int^1_{-1} d\mu~&e^{iA\mu}~\exp(B\mu^2)=2\sum_{n=0}^\infty\frac{(2n)!}{2^n n!}B^n\nonumber\\
&\times\sum_{p=0}^n(-2)^p\frac{4p+1}{(n-p)!(2n+2p+1)!!}j_{2p}(A),
\end{align}
where we have used the properties of $b_n^l$ coefficients and orthogonality of Legendre polynomials 
\begin{align}
\int_{-1}^1d\mu~P_{l_1}(\mu)P_{l_2}(\mu)=\frac{2}{2l_1+1}\df^D_{l_1l_2}.
\end{align}
Finally for the integral in equation \eqref{eq:epsint} it follows
\begin{align}
\int^1_{-1} & d\mu~ e^{iA\mu}~\exp(B\mu^2+i\epsilon\mu^3)\nonumber\\
&=2\sum_{p_1=0}^\infty (-)^{p_1}\Big((4p_1+1)F_{1}(p1,B,\epsilon)j_{2p_1}(A)\nonumber\\
&\qquad+(4p_1+3)F_{2}(p1,B,\epsilon)j_{2p_1+1}(A)\Big)
\label{eq:epsint2}
\end{align}
where we define the functions
\begin{align}
F_1(p1,B,\epsilon)&=\sum_{n=0}^\infty\sum_{r=0}^\infty c_1(p_1,n,r)B^n\epsilon^{2r},\nonumber\\
F_2(p1,B,\epsilon)&=\sum_{n=0}^\infty\sum_{r=0}^\infty c_2(p_1,n,r)B^n\epsilon^{2r+1},
\end{align}
and the coefficients are given by
\begin{widetext}
\begin{align}
c_1(p_1,n,r)&=\frac{(-)^{r}}{2^{n+3r}}\frac{(2n)!(6r)!}{n!(2r)!}
\sum_{p_2=0}^n \frac{2^{p_2}(4p_2+1)}{(n-p_2)!(2n+2p_2+1)!!}
\sum_{p_3=\left|p_1-p_2\right|}^{\text{min}\left\lbrace p_1+p_2,3r\right\rbrace} 
\frac{2^{p_3}\left\langle 2p_1,0,2p_2,0|2p_3,0\right\rangle^2}{(3r-p_3)!(6r+2p_3+1)!!},\nonumber\\
c_2(p_1,n,r)&=\frac{(-)^{r+1}}{2^{n+3r+1}}\frac{(2n)!(6r+3)!}{n!(2r+1)!}
\sum_{p_2=0}^n \frac{2^{p_2}(4p_2+1)}{(n-p_2)!(2n+2p_2+1)!!}\nonumber\\
&\qquad\qquad\qquad\qquad\qquad\qquad\qquad\times
\sum_{p_3=\frac{1}{2}\left(\left|2(p_1-p_2)+1\right|-1\right)}^{\text{min}\left\lbrace p_1+p_2,3r+1\right\rbrace} 
\frac{2^{p_3}\left\langle 2p_1+1,0,2p_2,0|2p_3+1,0\right\rangle^2}{(3r-p_3+1)!(6r+2p_3+5)!!}.
\end{align}
\end{widetext}
Here we have used the properties of $b_n^l$ coefficients, and introduced the Clebsch-Gordan coefficients
$\left\langle l_1,m_1,l_2,m_2|L,M\right\rangle$, which appeared as a solution of the integral over three 
Legendre polynomial
\begin{align}
\int_{-1}^1d\mu~P_{l_1}(\mu)P_{l_2}(\mu)P_{l_3}(\mu)=\frac{2}{2l_3+1}\left\langle l_1,0,l_2,0|l_3,0\right\rangle^2.
\end{align}
We stress that $c_1$ and $c_2$ are coefficients and do not depend on the values of $A$, $B$ or $\epsilon$. Moreover,
the sums that they contain are finite, thus, these coefficients are finite numbers themselves and can be precomputed. In
this way, they do not pose any computational obstacle, even though the expressions look somewhat formidable.
Also note that in this expansion $A$ term appears only as an argument of the spherical Bessel functions and the 
rest of the expansion is in powers of $B$ and $\epsilon$. 
In comparison with the method I in section \ref{sec:int1} we note that the advantage of method I is in partially resumming
the contribution of term $B$. In our case $A$, $B$ and $\epsilon$ are given by relations \eqref{eq:ABee}, and given the range 
of values for variables $q$, $k$, and terms $X$, $Y$, $V$ and $T$ (see figure \ref{fig:XYTV}) resummation of $B$ gives method 
I certain computational advantage. Finally, once convergence is reached methods I and II give the same results, as can be easily 
checked by comparison with the direct numerical evaluation of integrals on the l.h.s. of equations \eqref{eq:epsint} or 
\eqref{eq:epsint2} for some arbitrary real values of $A$, $B$ and $\epsilon$.

We also address the question of the difference of full result for terms $\mathcal{V}$ and $\mathcal{T}$ (in equation \eqref{eq:ZVT}) 
obtained using method I or/and II from the linearized version. In the linearized versions of integrals $\mathcal{V}$ and $\mathcal{T}$
only the leading order in terms $V$ and $T$ are kept. 
In figure \ref{fig:VT} we show the difference between the full and linearized result. We also show this difference relative to the 
two point power spectrum (labeled $\mathcal{Z}$ in equation \eqref{eq:ZVT}). As expected, the difference starts to appear 
at higher values of $k$, but since the total result is suppressed relative to the leading contribution it is hard to distinguish it 
on the total power spectrum, as we have mentioned earlier (see subsection \ref{subsec:angmom} and figure \ref{fig:PowerSpectrum}).

\vfill
\bibliographystyle{arxiv_physrev}
\bibliography{zel_pt}

\end{document}